\documentclass[iop]{emulateapj}

\usepackage{natbib}
\usepackage{amsmath}
\usepackage{amsfonts}
\usepackage{hyperref}

\usepackage{color}

\renewcommand{\vec}[1]{\mathbf{#1}}

\begin{document}

\title{Numerical simulations of quiet Sun magnetism: On the contribution from a
small-scale dynamo}
\shorttitle{}

\author{M. Rempel\altaffilmark{1}}

\shortauthors{Rempel}
 
\altaffiltext{1}{High Altitude Observatory,
    NCAR, P.O. Box 3000, Boulder, Colorado 80307, USA}

\email{rempel@ucar.edu}

\begin{abstract}
We present a series of radiative MHD simulations addressing the origin and distribution of mixed polarity
magnetic field in the solar photosphere. To this end we consider numerical simulations that cover the uppermost
$2-6$~Mm of the solar convection zone and we explore scales ranging
from $2$~km to $25$~Mm. We study how the strength and distribution of magnetic field
in the photosphere and subsurface layers depend on resolution, domain size and boundary conditions. We find that $50\%$
of the magnetic energy at the $\tau=1$ level comes from field with the less than $500$~G strength and that $50\%$ of
the energy resides on scales smaller than about $100$~km. 
While probability distribution functions are essentially independent
of resolution, properly describing the spectral energy distribution requires grid spacings of $8$~km or smaller. The formation of flux 
concentrations in the photosphere exceeding $1$~kG requires a mean vertical field strength greater than $30-40$~G at $\tau=1$. 
The filling factor of kG flux concentrations increases with overall domain size as magnetic field becomes organized by larger, 
longer lived flow structures. A solution with a mean vertical field strength of around $85$~G at $\tau=1$ requires a subsurface RMS 
field strength increasing with depth at the same rate as the equipartition field strength. We consider this an upper limit for the quiet Sun 
field strength, which implies that most of the convection zone is magnetized close to equipartition.
We discuss these findings in view of recent high-resolution spectropolarimetric observations of quiet Sun magnetism.
\end{abstract}


\received{}
\accepted{}

\maketitle
\section{Introduction}
\label{sect:intro}
Small-scale turbulent magnetic field is ubiquitous on the solar surface and provides the dominant contribution to the 
magnetic energy in the quiet Sun photosphere (see, e.g., \citet{Lites:1996:small_scale_field,
Khomenko:2003:quiet_sun_infrared,Dominguez:2003:quiet_sun,Almeida:2003:internetwork_sol_min,Trujillo:etal:2004,
Dominguez:2006:quiet_sun,Orozco-Suarez:etal:2007,Lites:etal:2008,Bellot:2012:pervasive_lin_pol} and recent reviews by
\citet{deWijn:2009:SSRv,Mpillet:2013:SSRv}). Several investigations
found that inter-network magnetic field shows only little dependence on the solar cycle \citep{Trujillo:etal:2004,Buehler:2013:cyc_dep} and also little correlation 
with the strength of the surrounding network field \citep{Lites:2011:hinode_ssd,Ishikawa:2009:cmp_QS_plage}. This points toward
an origin of the quiet Sun magnetic field largely independent from the global solar dynamo responsible for the solar
cycle.

It was suggested by \citet{Petrovay:1993:SSD}, based on a simplified transport model for signed and unsigned flux in the convection zone,
that a small-scale dynamo is the key process maintaining turbulent magnetic field in the quiet Sun. Small-scale dynamos were first 
studied through MHD simulations in incompressible setups by \citet{Cattaneo:1999} and later with stratification (anelastic approximation) 
by \citet{Bercik:2005:dynamo}. \citet{Voegler:Schuessler:2007} used the most "solar-like" setup by including realistic physics in terms 
of equation of state and 3-dimensional radiative transfer. They were able to demonstrate that despite the lack of significant recirculation 
within the computational domain (use of open bottom boundary conditions that mimic the deep convection zone), a considerable amount 
of magnetic field can be maintained in the photosphere. It was found later that the photospheric field strength falls still short by a 
factor of $2-3$ compared to observations based on Zeeman diagnostics \citep{Danilovic:2010:zeeman_dynamo}; an even more dramatic 
shortfall by about one order of magnitude was found by \citet{Shchukina:2011:hanle_dynamo} based on Hanle-effect diagnostics. To 
which degree the discrepancy between the magnetic field strength found in solar photospheric dynamo simulations and observations is 
due to boundary conditions and resolution (which relates to the magnetic Reynolds numbers reached) is still an open issue, which we aim 
to adress primarily in this paper.

On a more fundamental level the operation as well as non-linear saturation of small-scale dynamos at very small magnetic Prandtl 
numbers ($P_M=\nu/\eta \ll 1$, with viscosity $\nu$ and magnetic diffusivity $\eta$) as encountered on the Sun ($P_M\sim 10^{-5}$) 
remains an open question. While the regime $P_M > 1$ 
has been studied in great detail including the non-linear saturation phase \citep{Schekochihin:2004:SSD_high_Pm}, the regime $P_M\ll 1$ 
is less accessible by direct numerical simulations. Small-scale dynamos at low $P_M$ were studied in the kinematic regime by 
\citet{Iskakov:2007:SSD_low_Pm,Schekochihin:2007:SSD_low_PM} and in the non-linear regime by \citet{Brandenburg:2011:SSD_low_Pm}.
These investigations indicate that the threshold for dynamo action increases moderately when approaching the $P_M\ll 1$ regime, but
small-scale dynamo action remains possible for large values of the magnetic Reynolds number typically found in astrophysical
systems \citep{Tobias:Cattaneao:Boldyrev:SSD_review,Brandenburg:2011:SSD_low_Pm}.  

We present here an investigation which follows along the lines of comprehensive photospheric MHD simulations similar to the work presented by \citet{Voegler:Schuessler:2007,Pietarila-Graham:etal:2010:SSD}. In particular we address the problem that these simulations fall short by
a factor of about $2-3$ in field strength when compared to observations as pointed out by \citet{Danilovic:2010:zeeman_dynamo}. We explore
here the potential role of two factors: 1. Numerical resolution and magnetic diffusivities, 2. Influence from the bottom boundary condition.
While numerical resolution and treatment of diffusivities determine the dynamo growth rate in the kinematic phase, $\gamma_K$, the bottom boundary
condition determines the amount of magnetic energy that recirculates within the computational domain. It was first pointed out by 
\citet{Stein:2003:recirculation} that the rather small recirculation of plasma found in the top layers of the convection zone could be a major
hurdle for a small-scale dynamo to exist locally in the photosphere. While the recirculation is small it is not zero, since
some level of turbulent mixing between up- and downflow regions is unavoidable. Magnetic energy loss due to overturning convection happens on
a rather slow time-scale $\sim H_{\varrho}/v_{z {\rm RMS}}$ (with density scale height $H_{\varrho}$ and vertical RMS velocity $v_{z {\rm RMS}}$), 
which can be compensated by a sufficiently efficient small-scale dynamo. \citet{Voegler:Schuessler:2007} showed that a local
dynamo in the photosphere can operate despite small recirculation if the value of the magnetic diffusivity $\eta$ is smaller than 
$2.5\cdot 10^{10}\mbox{cm}^2\mbox{s}^{-1}$ (for a magnetic Prandtl number close to unity). 
\citet{Pietarila-Graham:etal:2010:SSD} studied photospheric dynamos in the kinematic growth phase using values of $\eta$ as low as
 $4\cdot 10^{9}\mbox{cm}^2\mbox{s}^{-1}$, but did not study the non-linear saturation. The setup of 
\citet{Voegler:Schuessler:2007,Pietarila-Graham:etal:2010:SSD} is conservative in terms of the bottom boundary condition. There is no vertical 
Poynting flux in upflow regions and in addition an enhanced magnetic diffusivity near the bottom boundary. Since it is likely that the bulk of the
convection zone has strong magnetic field, it is reasonable to assume that upflow regions are magnetized and transport magnetic energy into the
photosphere. 

In order to adress these questions we consider models that differ from previous studies in the following aspects:\vspace{0.1cm}\\1. We use only numerical diffusivities in an
attempt to minimize the influence from dissipation for a given numerical resolution. In comparison to direct numerical simulations (DNS) that use 
explicit diffusivities and fully resolve the dissipation range, our setup is more along the lines of large eddy simulations (LES) in which a high Reynolds number
regime is realized on large scales, while the dissipation range is truncated through the use of a subgrid-scale model. In our case the latter is entirely based
on monotonicity constraints through the use of a slope-limited diffusion scheme (see Section \ref{sect:setup} for detail).
This is an attempt to make the small-scale dynamo maximally efficient, i.e. to maximize the kinematic growth rate $\gamma_K$ for a given numerical resolution. 
This setup allows us to study a regime with strong non-linear feedback (saturated phase of the dynamo), but it does not allow us to address questions related 
to the magnetic Prandtl number (implicitly set by the numerical dissipation terms, in general close to unity). \vspace{0.1cm}\\2. We use generalized open boundary conditions
which allow also for the presence of (mixed polarity) horizontal field in upflow regions. These are setups that explore a stronger coupling between the top
layers (including photosphere) and the bulk of the convection zone. Formally these boundary conditions lead to different magnetic energy loss rates at the bottom boundary. While they do not strongly affect the kinematic growth phase of an efficient dynamo with $\gamma_K \gg v_{z {\rm RMS}}/H_{\varrho}$, they do become relevant in the non-linear
saturation regime, which we primarily focus on here. \vspace{0.1cm}\\ 
In addition we compare models with different resolutions as well as domain sizes to
evaluate the robustness of results.
Our aim is to investigate with this setup a small-scale dynamo operating in a regime that is consistent with observational constraints
on the quiet Sun magnetic field strength, like those inferred by \citet{Danilovic:2010:zeeman_dynamo}. This regime is currently not 
accessible with comprehensive solar MHD simulations that use only physical diffusion terms (DNS) or even operate in a low $P_M$ regime. 

The remainder of the paper is organized as follows: In Section \ref{sect:setup} we describe in detail the numerical setup in terms of the equations
solved, the formulation of numerical diffusivites, boundary conditions and domain sizes used. In Section \ref{sect:results} we present
the results, subsections describe in detail the resolution dependence of results, the dependence on domain size and boundary conditions, and a
detailed analysis of the dynamo process based on transfer functions in spectral space. In Section \ref{sec:discussion} we discuss our main findings
in relation to observational constraints on quiet Sun magnetism, a detailed comparison with observations through forward modeling of spectral lines 
is deferred to future publications. Concluding remarks are presented in Section \ref{sect:conclusion}.

\section{Numerical setup}
\label{sect:setup}
\subsection{Numerical scheme}
\label{sect:numdiff}
We use for our simulations the {\em MURaM} radiative MHD code
\citep{Voegler:etal:2005,Rempel:etal:2009}. This code uses a $4^{th}$
order accurate (in space and time) conservative, centered finite 
difference scheme for discretization of the MHD equations, combined
with a short characteristics approach for radiative transfer.  The code uses a tabulated
{\em OPAL} equation of state \citep{Rogers:opal:1996}. We solve the MHD equations in the following form:
\begin{eqnarray}
\frac{\partial \varrho}{\partial t}&=&-\nabla\cdot\left(\varrho \vec{v}\right)\label{eq:cont}\\
\frac{\partial \varrho\vec{v}}{\partial t}&=&-\nabla\cdot(\varrho\vec{v}\vec{v}) +\frac{f_{v_A}}{4\pi}\nabla\cdot\left(\vec{B}\vec{B}-\frac{1}{2}\vec{I}\,B^2\right)\nonumber\\
&&-\nabla P +\varrho\vec{g}\label{eq:mom}\\
\frac{\partial E_{\rm HD}}{\partial t}&=&-\nabla\cdot\left[\vec{v}\,(E_{\rm HD} + P)\right]+\varrho\vec{v}\cdot\vec{g}+\frac{\eta}{4\pi}(\nabla\times\vec{B})^2 \nonumber\\
&&+\vec{v}\cdot \frac{f_{v_A}}{4\pi}\nabla\cdot\left(\vec{B}\vec{B}-\frac{1}{2}\vec{I}\,B^2\right)+Q_{\rm rad}\label{eq:ener}\\
\frac{\partial \vec{B}}{\partial t}&=&\nabla\times\left(\vec{v}\times\vec{B}-\eta\nabla\times\vec{B}\right)\label{eq:ind}
\end{eqnarray}  
Here $\varrho$, $P$, $\vec{v}$, and $\vec{B}$ denote mass density, pressure, velocity and magnetic field. For the
gravitational acceleration $\vec{g}$ we use a constant value of $-2.74\cdot 10^4 \mbox{cm}^2\mbox{s}^{-1}$ in the vertical direction (small local domains),
$\eta$ is an optional magnetic diffusivity. Furthermore $Q_{\rm rad}$ denotes the radiative heating term. We use
in the energy equation Eq. \ref{eq:ener} a treatment that is conservative for the quantity $E_{\rm HD}=E_{\rm int}+0.5\varrho v^2$ 
($E_{\rm int}$ denotes the internal energy). We separated out magnetic energy to avoid numerical problems in regions with
small values of the plasma beta $8\pi P/B^2$, which can be encountered above the photosphere in strong field regions. 
In addition the Lorentz force pre-factor $f_{v_A}$ can be used to artificially limit the Alfv\'en velocity in those regions in order to
prevent severe time step constraints. We use here the same functional form as \citet{Rempel:etal:2009}
\begin{equation}
f_{v_A}=\frac{1}{\sqrt{1+\left(\frac{v_A}{v_{\rm max}}\right)^4}}\;,
\end{equation}
where $v_A=B/\sqrt{4\pi\varrho}$ and $v_{\rm max}$ denotes the maximum permissible Alfv\'en velocity. 
While the latter two features were mostly implemented for sunspot simulations,
we limit also here in all simulations the maximum Alfv\'en velocity to $31.6\,\mbox{km s}^{-1}$ in order to prevent severe numerical time step constraints 
that can arise from strong magnetic field near the top boundary (in particular in the two simulations that extend $1.5$~Mm above the photosphere).
This does not impact any of the results presented here for which mostly the sub-photospheric dynamics matter ($f_{v_A}=1$ until about a few $100$ km above the
photosphere), but it dramatically reduces the required 
computing time by more than a factor of $10$ in some cases. This allows us to focus our study on higher resolution setups.
We do not consider explicit viscosity and also set $\eta$ to zero except for one control experiment. As a
consequence we require additional artificial diffusion terms in order to maintain numerical stability. These terms are
computed as described below.

We use here a modified version of the scheme first introduced
by \citet{Rempel:etal:2009}, which we explain here in detail. Our 
approach is based on a slope-limited diffusion scheme that uses a
piecewise linear reconstruction of the discrete solution $u_i$ to compute 
extrapolated values at cell interfaces:
\begin{eqnarray}
u_l&=&u_i+0.5\,\Delta u_i\\
u_r&=&u_{i+1}-0.5\,\Delta u_{i+1}\;.
\end{eqnarray}
Here $\Delta u_i$ denotes the reconstruction slope for the $i^{th}$ cell, 
$u_l$ ($u_r$) are the interface values extrapolated from the cells on the 
left (right). The reconstruction slopes $\Delta u_i$ are computed using the
monotonized central difference limiter, given by
\begin{eqnarray}
  \Delta u_i&=&\mbox{minmod}\left[(u_{i+1}-u_{i-1})/2,\right.\\ \nonumber
    &&\left. 2\,(u_{i+1}-u_i),2\,(u_i-u_{i-1})\right]\;.
\end{eqnarray}
Numerical diffusive fluxes at cell interfaces are computed from the
extrapolated values through the expression
\begin{equation}
  f_{i+\frac{1}{2}}=-\frac{1}{2}\,c_{i+\frac{1}{2}}\,
  \Phi_h(u_r-u_l,u_{i+1}-u_i)\cdot(u_r-u_l)\;.
\end{equation}
Here $c_{i+\frac{1}{2}}$ is a characteristic velocity at the cell interface,
the function $\Phi_h$ is given by
\begin{equation}
  \Phi_h=\mbox{max}\left[0,1+h\left(\frac{u_r-u_l}{u_{i+1}-u_i}-1\right)\right]
  \label{phi_lim}
\end{equation}
in regions with $(u_r-u_l)\cdot(u_{i+1}-u_{i})>0$, while $\Phi_h=0$ if
$(u_r-u_l)\cdot(u_{i+1}-u_{i})\leq 0$ (no anti-diffusion). Here $h$ is
a parameter that allows to control the (hyper-) diffusive character of the
scheme. A choice of $h=0$ reduces the diffusive flux to that of a standard 
second order Lax-Friedrichs scheme. For $h>0$ the diffusivity is reduced
for smooth regions in which $\vert(u_r-u_l)/(u_{i+1}-u_{i})\vert<1$, while
the maximum diffusivity of $0.5\,c_{i+\frac{1}{2}}\,\Delta x$ is always kept
in regions with $\vert(u_r-u_l)/(u_{i+1}-u_{i})\vert=1$. For values of
$h>1$ the diffusive fluxes are switched off in regions with
$\vert(u_r-u_l)/(u_{i+1}-u_{i})\vert<1-1/h$, leading to a diffusivity that
is concentrated to monotonicity changes or features resolved by only a few
grid points. For the work presented here we use a choice of $h=2$. 
\citet{Rempel:etal:2009} used a different functional form of 
$\Phi_h=[(u_r-u_l)/(u_{i+1}-u_{i})]^2$ in regions with 
$(u_r-u_l)\cdot(u_{i+1}-u_{i})>0$ that suppresses, but does not completely
disable diffusion for well-resolved features.

The above describe diffusion scheme is applied to the variables
$\{\log(\varrho), v_x, v_y, v_z, \varepsilon, B_x, B_y, B_z\}$, where
$\varepsilon=E_{\rm int}/\varrho$. In addition we make the assumption that
the diffusive mass flux also transports momentum and internal energy, i.e.,
we add to the momentum flux a term $\vec{f_{\varrho}}\vec{v}$ and to the 
energy flux a term $\vec{f_{\varrho}}\varepsilon$, where $\vec{f_{\varrho}}$ 
denotes
the diffusive mass flux. This correction is identical with the assumption 
that momentum and energy are transported by the total mass flux 
$\varrho\vec{v}+\vec{f_{\varrho}}$. Since at the same time the induction
equation uses only the velocity $\vec{v}$ without a contribution from the
diffusive mass flux, the presence of mass diffusion mimics to some degree
ambipolar diffusion.

For enhanced stability we also implemented a switch, which limits the maximum
density contrast between neighboring grid cells to $10$. If the density
contrast exceeds that threshold we disable the piecewise linear reconstruction
and set the diffusivity to the maximum value allowed for by the CFL condition
to prevent a further increase.

We also added an additional optional $4^{th}$ hyper-diffusion term that scales 
with the advection velocity and acts only in the vertical direction on the 
quantities $\log(\varrho)$, $v_z$, and $\varepsilon$. This term allows to damp
some low level spurious oscillations on the grid scale that are too small to
cause monotonicity changes in the presence of a background gradient 
(stratification) and go mostly undetected by the slope-limited diffusion scheme.

The numerical diffusion scheme is implemented in a dimensional split way to 
ensure maximum stability and is applied to the solution in a separate filtering
step after a full time-step update of our $4^{th}$-order time integration 
scheme. In the energy equation we account for artificial viscous and ohmic
heating.

Errors caused in ${\rm div}\vec{B}$ are controlled with the help of an iterative
hyperbolic divergence cleaning approach \citep{Dedner:etal:2002:divB}.

Estimating the effective diffusivity of our numerical scheme is not a trivial task. The numerical diffusivity 
is in general highly intermittent and inhomogeneous as well as scale-dependent (see Section \ref{sec:transfer} for
further detail). Comparing results obtained at $4$~km grid spacing with simulations that use only a physical 
magnetic diffusivity of $\eta=5\cdot 10^{9}\mbox{cm}^2\mbox{s}^{-1}$ (which is the minimum value
required for numerical stability in that case) we find an about $6$ times larger kinematic growth rate 
with numerical diffusivity, indicating a significantly lower effective diffusivity.

\subsection{Domain size, boundary conditions, simulation setup}
We present numerical simulations in two domains: 
$6.144\times 6.144\times 3.072\,\mbox{Mm}^3$ and $24.576\times
24.576\times 7.680\,\mbox{Mm}^3$. In the smaller domain the top boundary condition 
is located about $700$~km above the average $\tau=1$ level, in the large domain about 
$1.5$~Mm. This leads to depths of the convective part of about $2.3$~Mm and $6.2$~Mm, 
respectively. In addition we performed also a series of simulations in a
$98.304\times 98.304\times 18.432\,\mbox{Mm}^3$ sized domain, but we will not
discuss them in great detail in this publication.

All simulations presented here use a setup with no vertical netflux, i.e.
$\langle B_z\rangle=0$. Since we use for most setups open boundary conditions
and allow for the transport of horizontal flux across the bottom boundary, the domain 
averaged horizontal flux
can fluctuate, but stays on average close to zero. Our primary aim is to study the
contributions from a small-scale dynamo to quiet Sun magnetism separate from 
potential contributions of a large-scale dynamo. We will discuss how both
dynamos could be coupled in Section \ref{sect:discuss_LSD-SSD}.

In the horizontal direction the domains are periodic, the top boundary is
semi-transparent (open for outflows, closed for inflows). For the magnetic
field we use two top boundary conditions: vertical magnetic field and a potential 
field extrapolation.

Since the details of the formulation of the bottom boundary condition
have significant influence on the solutions in terms of the saturation field
strength reached, we explore here a total of $5$ different boundary
conditions. These boundary conditions are a balance between a self-contained
dynamo problem (best achieved with closed boundaries) and an attempt to
capture the deep convection zone (open boundaries).

In our numerical formulation
we have 2 ghost cells and the position of the domain boundary is between the first domain
and first ghost cell. For many variables we use boundary conditions which prescribe a
symmetric or anti-symmetric behavior across the boundary. If $v_1$ and $v_2$ are the
values in the first and second domain cell and $v_1^*$ and $v_2^*$ are the corresponding
quantities in the first and second ghost cell ($v_1^*$ is the ghost cell closest to the boundary),
a symmetric boundary implies $v_1^*=v_1$ and $v_2^*=v_2$, an anti-symmetric boundary implies
$v_1^*=-v_1$ and $v_2^*=-v_2$. 

Most of our simulations use open hydrodynamic boundary conditions, which aim
to mimic the presence of a deep convection zone beneath the domain boundary. We use here 
two different formulations for open and one formulation for a closed boundary condition, 
which we describe first before we detail the magnetic boundary conditions:
\begin{enumerate}
  \item[\it HD1:]
    All three mass flux components are symmetric with respect to the boundary. The pressure 
    $P_{\rm BND}=P_{\rm gas}+B_z^2/(8\pi)$ is uniform and fixed at the boundary.
    If $P_1$ and $P_2$ are the values of the gas pressure in the first and second domain cell, we assign 
    the ghost cell values as follows (linear extrapolation into ghost cells):
    \begin{eqnarray}
      P_1^*&=&1.5\,P_{\rm gas}-0.5\,\sqrt{P_1\,P_2}\\
      P_2^*&=&2.5\,P_{\rm gas}-1.5\,\sqrt{P_1\,P_2}
    \end{eqnarray}
    The entropy is symmetric in downflow regions and is specified in upflow regions such that the 
    resulting radiative losses in the photosphere lead to a solar-like energy flux (within a few $\%$).
    The corresponding values for density and internal energy follow from the equation of state.
    In addition upflow velocities are capped at $1.5$ times the vertical RMS velocity at the boundary to 
    prevent extreme events.
  \item[\it HD2:]
    All three mass flux components are symmetric with respect to the boundary. We decompose the gas pressure 
    into mean pressure and fluctuation, $P=\bar{P}+P^{\prime}$. The mean pressure is extrapolated into the ghost cells such that its
    value at the boundary is fixed, while the pressure fluctuations are damped in the ghost cells. This is achieved the following way:
    \begin{eqnarray}
      \bar{P}_1^*&=&\bar{P}_1 \cdot\frac{P_{\rm BND}}{\sqrt{\bar{P}_1 \bar{P}_2}}\\
      \bar{P}_2^*&=&\bar{P}_1 \cdot\frac{P_{\rm BND}^2}{\bar{P}_1 \bar{P}_2}\\
      P_1^{\prime *}&=&P_1^{\prime}\cdot C_{\rm dmp}\label{p1_dmp}\\
      P_2^{\prime *}&=&P_1^{\prime}\cdot C_{\rm dmp}^2\label{p2_dmp}\    
    \end{eqnarray}
    We use a value of $C_{\rm dmp}=0.95$. We used first a symmetric boundary condition for $P^{\prime}$, but found
    problems with over-excited standing pressure waves in deeper domains. 
    The entropy is symmetric in downflow regions and is specified in upflow regions such that the 
    resulting radiative losses in the photosphere lead to a solar-like energy flux (within a few $\%$).
    The corresponding values for density and internal energy follow from the equation of state.
  \item[\it HD3:]
    This is a closed boundary condition.   The vertical mass flux is antisymmetric, the horizontal  velocity components are 
    symmetric (closed for vertical mass flux and stress free for horizontal motions).
    The gas pressure is extrapolated into the ghost cells as follows:
   \begin{eqnarray}
      P_1^*&=&P_1^2/P_2\\
      P_2^*&=&P_1^3/P_2^2\,.
    \end{eqnarray}   
    The entropy is symmetric across the boundary. We added a heating term in the lower
    $10\%$ of the domain to replenish the energy radiated away in the photosphere.
\end{enumerate}
We used in our investigation initially the boundary {\it HD1}. Since the pressure at the boundary is fixed, this
boundary condition does not allow for pressure differences between up- and downflow regions, which are expected
for dynamical reasons.  As a consequence this boundary condition underestimates the value of horizontal flow divergence 
in upflow regions when compared to a deeper reference run. The boundary condition {\it HD2} puts less constraints on the
pressure at the boundary and does allow for systematic pressure differences between up- and downflow regions and improves
the properties of the flow at the boundary. While {\it HD1} accounts only for magnetic pressure from vertical field, {\it HD2} 
incorporates the total magnetic pressure to the degree it is reflected in the gas pressure perturbation $P^{\prime}$ (we exclude
magnetic pressure contributions from the damping in Eqs. \ref{p1_dmp} and \ref{p2_dmp}). 
The boundary  {\it HD3} is used for control experiments using a closed domain. 

In addition to the above described hydrodynamic boundary conditions we implement the following magnetic boundary conditions
in our experiments: 
\begin{enumerate}
  \item[\it OV:]
    (Open boundary/vertical field)
    We use {\it HD1}, the magnetic field is vertical at the boundary ($B_z$ symmetric, $B_x$ and $B_y$ antisymmetric).
  \item[\it OSa:]
    (Open boundary/symmetric field) 
    We use {\it HD1}, all three magnetic field components are symmetric. We impose an upper limit of
    $600$~G for the horizontal RMS field strength in inflow regions and limit the maximum horizontal magnetic field 
    strength to $3$ times the RMS value. We set net horizontal magnetic flux in inflow regions to zero and
    rescale the vertical magnetic field such that the horizontal and vertical RMS field strength are identical 
    in inflow regions (since we consider here only situations with $\langle B_z\rangle=0$ the rescaling
    of $B_z$ does not affect the vertical net flux).
  \item[\it OSb:]
    (Open boundary/symmetric field)
    We use {\it HD2}, all three magnetic field components are symmetric.
  \item[\it OZ:]
    (Open boundary/zero field)
    We use {\it HD2}, similar to {\it OSb}, but we set $\vec{B}=0$ in inflow regions, i.e. $\vec{B}$ is antisymmetric
    in inflow and symmetric in outflow regions.
  \item[\it CH:]
    (Closed boundary/horizontal field)
    We use {\it HD3}, the magnetic field is horizontal at the bottom boundary ($B_z$ antisymmetric, 
    $B_x$ and $B_y$ symmetric).
\end{enumerate}
The boundary condition {\it OV} is similar to that used by \citet{Voegler:Schuessler:2007} and
we included one simulation with this boundary condition to better connect our results to
previous work. We started our investigation with {\it OSa}, but found that we had to implement
several corrections to the magnetic field to prevent runaway solutions when we also allow for
a horizontal magnetic to be present in inflow regions in combination with the hydrodynamic boundary
condition {\it HD1}. Most importantly, we limit the horizontal
RMS field strength to $600$~G (for the $6.144\times 6.144\times 3.072\,\mbox{Mm}^3$ domain),
which corresponds to a solution in which $B_{\rm RMS}$ increases with depth approximately at the
same rate as the equipartition field strength $B_{\rm eq}=\sqrt{4\pi \varrho}\,v_{\rm RMS}$ 
(see Section \ref{sec:subsurf} for further detail). Using the hydrodynamic boundary condition {\it HD2}
resolves most of these issues and a much simpler magnetic boundary conditions is sufficient ({\it OSb}).
The differences between boundary conditions {\it OSa,b} affect mostly the first
pressure scale height above the bottom boundary, boundary {\it OSb}  performs overall
better when comparing simulations with different domain depths (see Section \ref{sec:subsurf}).
The boundary {\it OZ} is a control experiment making the very conservative (and likely
unrealistic) assumption that the deep convection zone is unmagnetized. We use boundary {\it CH}
as an additional control experiment to study a setup in which we have a complete recirculation
of mass and all magnetic induction effects are confined to the simulation domain.

As a general note we want to point out that none of the above boundary conditions is "perfect''.
Closed boundary are not a representation for the deep solar convection zone and open boundaries suffer
all from the same problem that the properties of quantities leaving the domain are well determined by
the solution, while the properties of quantities entering the domain have to be assumed, i.e. these
boundary conditions cannot be free from implicit or explicit assumptions.  It is therefore crucial
to compare simulations with different boundary conditions as well as domain depths in order to quantify their
potential influence on solution properties.

\begin{table}
  \begin{center}
    \begin{tabular}{ c | c | c | c | c }
      \hline
      ID & Size [Mm$^3$] & Res [km] & Bot & Top \\ \hline\hline
      V16  & $6.144\times 6.144\times 3.072$  & $16$ & {\it OV} & V\\ \hline\hline
      O32a & $6.144\times 6.144\times 3.072$  & $32$ & {\it OSa} & V\\ \hline
      O16a & $6.144\times 6.144\times 3.072$  & $16$ & {\it OSa} & V\\ \hline
      O8a & $6.144\times 6.144\times 3.072$  & $8$  & {\it OSa} & V\\ \hline
      O4a & $6.144\times 6.144\times 3.072$  & $4$  & {\it OSa} & V\\ \hline
      O2a & $6.144\times 6.144\times 3.072$  & $2$  & {\it OSa} & V\\ \hline\hline
      O32b & $6.144\times 6.144\times 3.072$  & $32$ & {\it OSb} & P\\ \hline
      O16b & $6.144\times 6.144\times 3.072$  & $16$ & {\it OSb} & P\\ \hline
      O8b & $6.144\times 6.144\times 3.072$  & $8$  & {\it OSb} & P\\ \hline\hline
      Z32 & $6.144\times 6.144\times 3.072$  & $32$ & {\it OZ}  & P\\ \hline 
      Z16 & $6.144\times 6.144\times 3.072$  & $16$ & {\it OZ}  & P\\ \hline
      Z8 & $6.144\times 6.144\times 3.072$  & $8$  & {\it OZ}  & P\\ \hline\hline
      C32 & $6.144\times 6.144\times 3.072$  & $32$ & {\it CH} & V\\ \hline
      C16 & $6.144\times 6.144\times 3.072$  & $16$ & {\it CH} & V\\ \hline     
      C8 & $6.144\times 6.144\times 3.072$  & $8$ & {\it CH}  & V\\ \hline
      C8$\eta$ & $6.144\times 6.144\times 3.072$  & $8$ & {\it CH}  & V\\ \hline \hline 
      O16bM & $24.576\times 24.576\times 7.68$ & $16$ & {\it OSb} & P\\ \hline
      Z16M & $24.576\times 24.576\times 7.68$ & $16$ & {\it OZ} & P\\ \hline  \hline
      O32bSG & $98.304\times 98.304\times 18.432$ & $32$ & {\it OSb} & P\\ \hline  \hline
    \end{tabular}
  \end{center}
  \caption{Overview of numerical simulations discussed in this publication. See text 
    for further explanation.}
  \label{tab:t1}
\end{table}

In Table \ref{tab:t1} we present all the simulations we discuss in this publication.
With the exception of O2a, C8, and C8$\eta$ all simulations were started from a thermally relaxed 
non-magnetic convection simulation after addition of a $\sim 10^{-3}$~G random
field (pointing in the z-direction, random in the horizontal plane and uniform in the vertical direction). 
The run O2a was restarted from a saturated snapshot of O4a and evolved for 
an additional 5 minutes to further explore the resolution dependence. The simulation
C8 and C8$\eta$ were restarted from C16. C8$\eta$ uses a Laplacian diffusivity of 
$\eta=10^{10}\,\mbox{cm}^2\,\mbox{s}^{-1}$ for the magnetic field instead of numerical
diffusivity (we kept numerical diffusivity for all other variables). 

The simulation O32bSG was restarted from a sequence of lower resolution runs we do not list in Table \ref{tab:t1}.
As a consequence the spectral energy distribution is in this run likely biased toward larger scales. We use
this simulation here mostly to explore the connection toward deeper layers of the convection zone through
comparison of horizontally averaged mean quantities.

The column "Bot" refers to the boundary condition used at the bottom boundary,
the column "Top" to the magnetic field boundary condition used at the top. Here
"V" and  "P" refer to vertical magnetic field and potential field extrapolation.
The hydrodynamical boundary condition at the top boundary is in all cases open for upflows (i.e. upward
directed shocks can leave the domain) and closed for downflows. In the simulations 
O16bM and Z16M the top boundary is about $1.5$~Mm above the photosphere, while it is about
$700$~km in all other simulations. 

\begin{figure*}
  \centering 
  \resizebox{0.85\hsize}{!}{\includegraphics{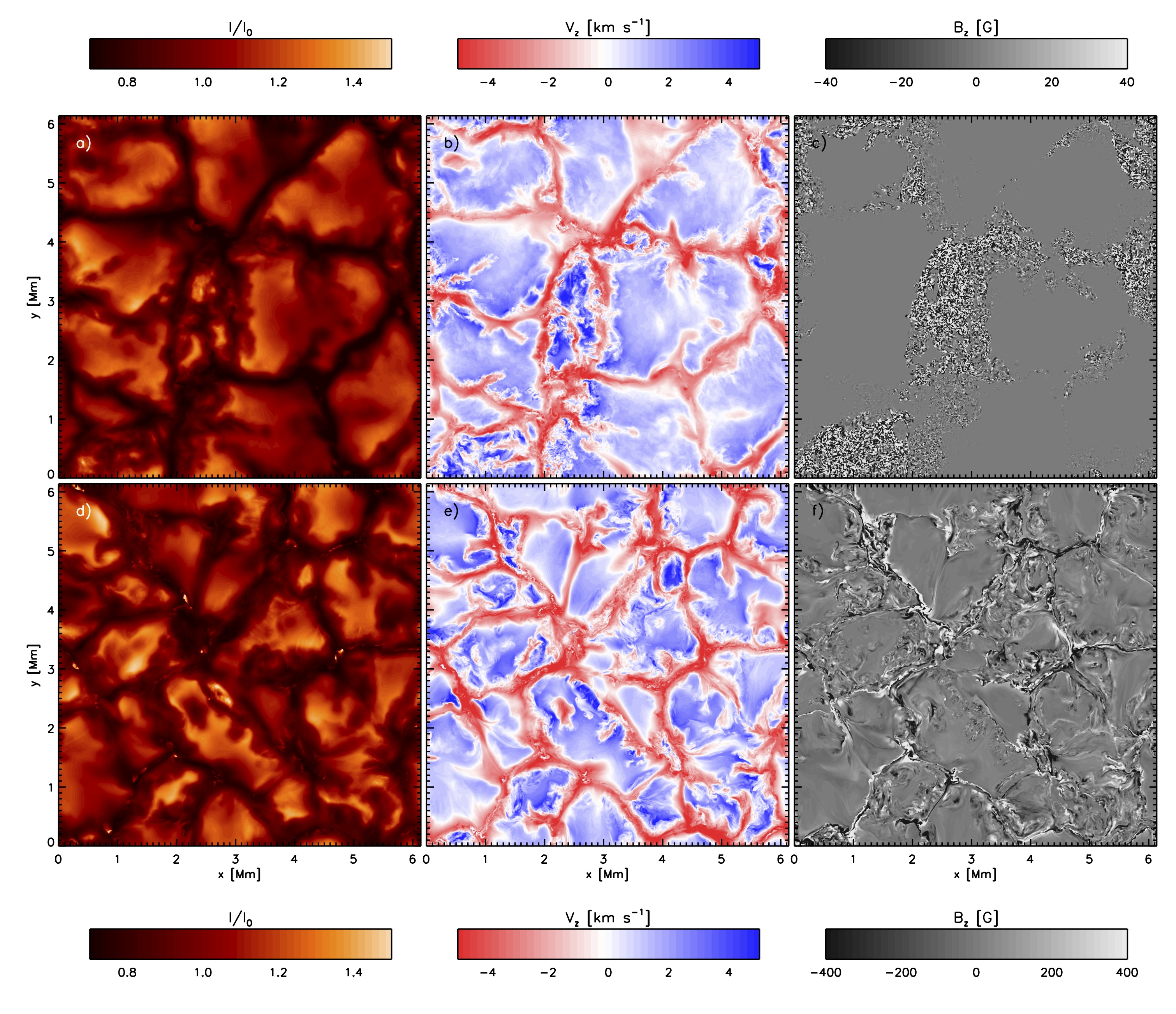}}
  \caption{Comparison of kinematic growth phase (panels a-c) and saturated 
    phase (panels d-f) of the simulation run O4a. Shown are the bolometric 
    intensity (panels a, d), the vertical velocity at $\tau=1$ (panels b, e), 
    and the magnetogram at $\tau=1$ (panels c, f). Two animations of this figure
    for the kinematic and saturated phase are provided with the online
    material. During the kinematic growth phase we clipped the displayed values
    of $B_z$ in the movie at $\pm 5 \langle \vert B_z\vert\rangle$ to follow the
    fast growth of the field; in the saturated phase we display values in the range
    $\pm 400$~G similar to panel f.
  }
  \label{fig:IVzBz_kin_sat}
\end{figure*}

\subsection{Scope of the simulations presented here}
Are the simulations we present here small-scale dynamos? This question arises because
of two aspects of our setup: open boundary conditions and the use of (unphysical)
numerical diffusivities. The open boundaries we use allow for a magnetic energy flux across domain
boundaries,
which implies that the maintenance of the magnetic field is not restricted to processes
within the simulation domain. Although, as we show later, the Poynting flux transports
significantly more energy out of the domain than is returning back in inflow regions.     
We have conducted experiments that use closed boundary conditions and only a physical
Laplacian diffusivity for the magnetic field (run C8$\eta$) and we confirmed that we have a small
scale dynamo operating under these conditions. In addition a comparison of the spectral
energy transfers presented in Section \ref{sec:transfer} does not reveal any significant differences
(apart from the saturation field strength reached) between this reference simulation and a simulation 
solely based on numerical diffusivities.
While our numerical experiments should be more carefully labeled as large eddy simulations of
photospheric magneto-convection with zero imposed magnetic flux, we did not find any indication
that they are not small-scale dynamos.

Since we apply the same numerical dissipation scheme to all MHD variables, the resulting
"numerical magnetic Prandtl number'' is close to 1 in all our experiments. We do not address here
the role of the magnetic Prandtl number for the small-scale dynamo process.  

\section{Results}
\label{sect:results}
In the following subsections we analyze our simulations by presenting quantities in the photosphere
on constant $\tau$ levels. Since we use here only simulations computed with gray radiative transfer
these layers refer to a $\tau$-scale computed with mean opacities. Further $\tau$ levels always refer 
to warped $\tau=\mbox{const.}$ surfaces and not the constant geometric height surface with the corresponding
average $\tau$ value.   

We further discuss in detail (mostly) photospheric power spectra and probability distribution functions
of magnetic field. On the one hand we use these quantities to simply compare different simulations, on the
other hand they have a strong connection to results from observational studies of quiet Sun magnetism.
We refer the reader to Sections \ref{sec:discuss_PDF} and \ref{sec:discuss_power} for a summary and discussion 
of their importance. 

\begin{figure*}[ht!]
  \centering 
  \resizebox{0.8\hsize}{!}{\includegraphics{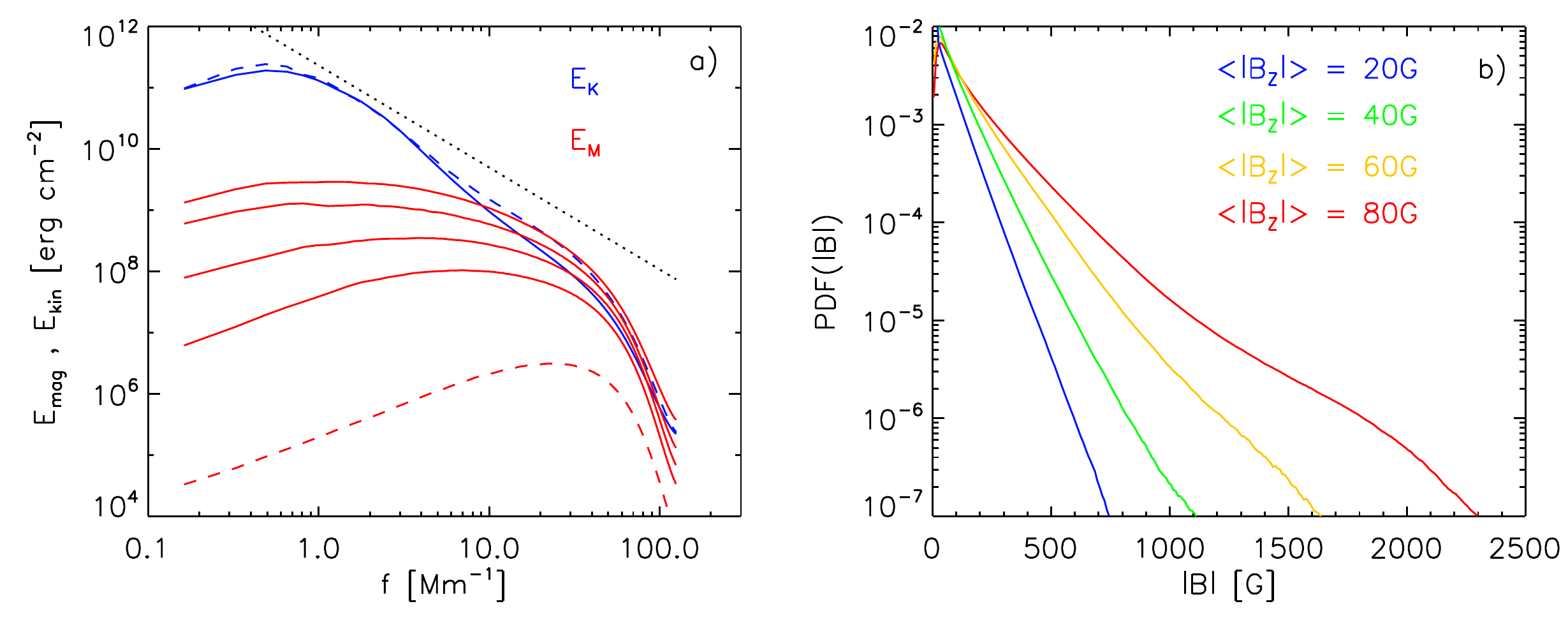}}
  \caption{a) Photospheric ($\tau=1$) power spectra for magnetic energy (red) 
    and kinetic energy (blue) in simulation O4a. The dashed lines correspond to the kinematic 
    growth phase during times when Lorentz-force feedback was negligible. We
    scaled up the magnetic energy spectrum in order to show clearly in the same figure.
    The solid red lines correspond bottom to top
    to snapshots with $20, 40, 60$, and $80$~G vertical mean field strength 
    at $\tau=1$. For increasing field strength the peak of magnetic power moves
    toward larger scales. The solid blue line shows the kinetic energy spectrum 
    for the $80$~G case. The dotted line in panel a) indicates a Kolmogorov slope
    of $-5/3$. b) Probability distribution functions for
    $\vert B\vert$ at $\tau=1$ for vertical mean field strength from $20$ to $80$~G. 
  }
  \label{fig:pwr_pdf_sat}
\end{figure*}

\begin{figure*}
  \centering 
  \resizebox{0.95\hsize}{!}{\includegraphics{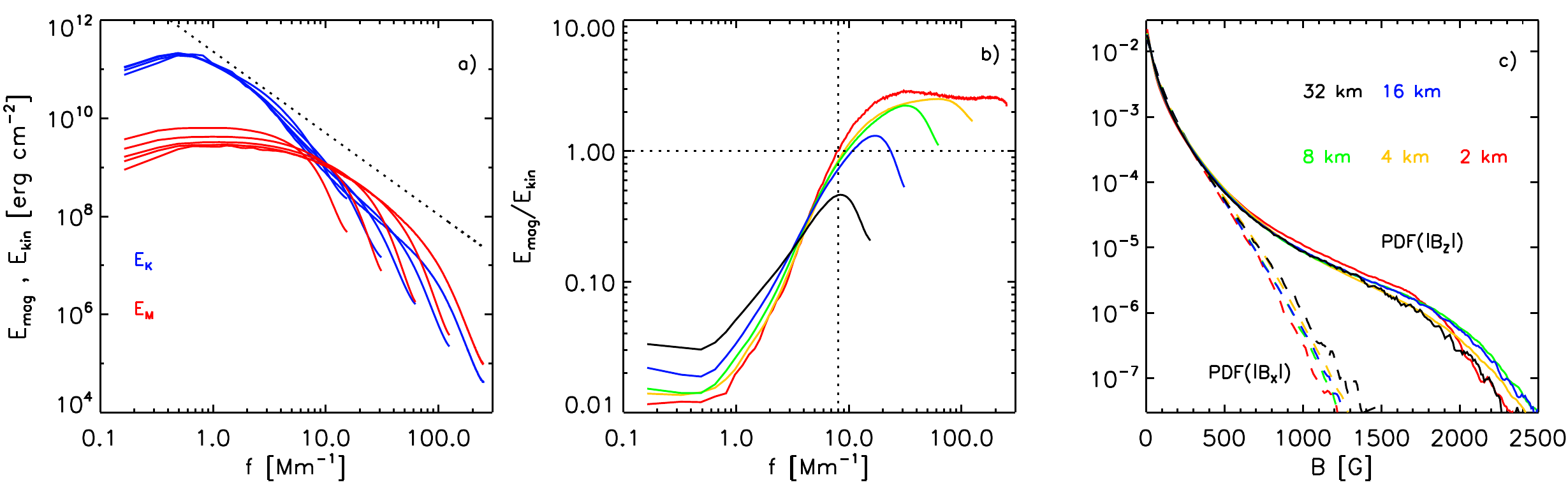}}
  \caption{a) Resolution dependence of magnetic (red) and kinetic (blue) energy
    power spectra comparing the simulations O32a-O2a. All simulations reached about
    $80$~G vertical mean field strength at $\tau=1$, the grid spacing was varied from $32$ to $2$~km. 
    The dotted line in panel a) indicates a Kolmogorov slope of $-5/3$.
    b) Ratio of magnetic to kinetic energy as function
    of resolution. For grid spacings smaller than $16$~km the magnetic energy is
    in super-equipartition on scales smaller than about $100$~km. For the highest
    resolution cases ($4$ and $2$~km) we see some indication that the ratio
    might reach asymptotically a value around  $2$. c) Resolution dependence
    of the probability distribution function for $\vert B_x\vert$ and $\vert B_z\vert$. 
  }
  \label{fig:pwr_pdf_res}
\end{figure*}

\begin{figure*}
  \centering 
  \resizebox{0.75\hsize}{!}{\includegraphics{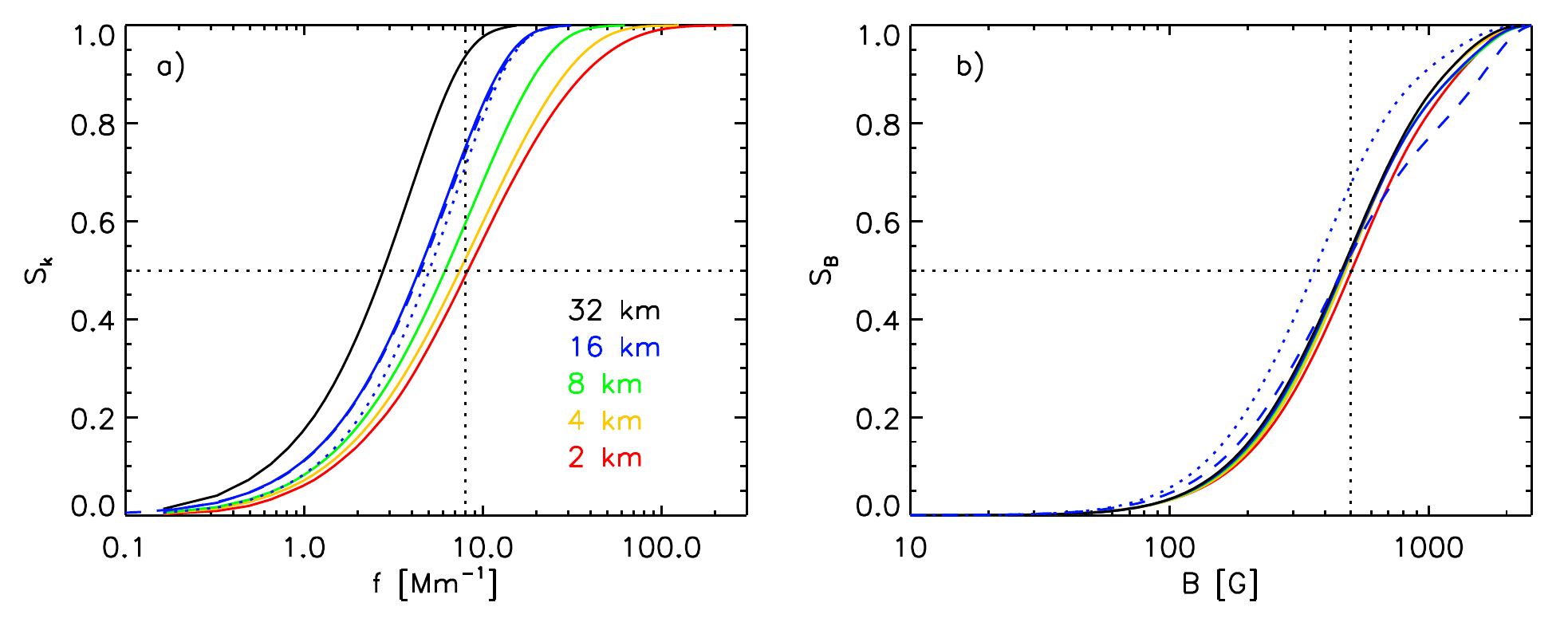}}
  \caption{a) Normalized integrated magnetic energy spectra from Figure \ref{fig:pwr_pdf_res}.
    For the highest resolution case (grid spacing of $2$~km) $50\%$ of the magnetic
    energy found at the $\tau=1$ surface is present on scales smaller than $100$~km.
    b) Normalized integrated magnetic energy distribution functions. $50\%$ of the
    magnetic energy on the $\tau=1$ surface is found in field with less than $500$~G
    strength. Solid lines correspond to simulations O32a-O2a with 
    $\langle\vert B_z\vert\rangle=80$~G, dashed (dotted) lines correspond to simulations
    O16b and O16bM with $\langle\vert B_z\vert\rangle=60$~G.
  }
  \label{fig:energy_dist}
\end{figure*}

\subsection{Kinematic to saturated phase}
\label{sec:kin_sat}
We start our discussion of results with numerical simulations using the 
$6.144\times 6.144\times 3.072\,\mbox{Mm}^3$ domain and the boundary condition {\it OSa}.
We limit the horizontal RMS field strength in inflow regions to $600$~G, which corresponds
approximately to a solution in which the RMS field strength increases with depth as the
same rate as the equipartition field strength. As we will discuss in Section \ref{sec:subsurf}, these
solutions are close to an upper limit for the quiet Sun field strength.
We use the small domain to explore the resolution dependence of the results and repeated the
same experiment with grid spacings of $32$, $16$, $8$, and $4$~km. All simulations were started 
from a thermally relaxed $B=0$ G convection simulation to which we added a $10^{-3}$~G seed field
(pointing in the z-direction, random in the horizontal plane and uniform in the vertical direction). In 
addition we present a simulation with $2$~km grid spacing, which was restarted from the $4$~km case.

Figure \ref{fig:IVzBz_kin_sat} presents for the simulation with $4$~km grid spacing (O4a) two snapshots, one
during the early growth phase (panels a-c) at a time when 
$\langle\vert B_z\vert\rangle(\tau=1) = 8.4$~G and one during during a later phase (panels d-f) when
$\langle\vert B_z\vert\rangle(\tau=1) = 86$~G. The panels a) and d) show the intensity
for a vertical ray, panels b) and e) the vertical velocity at $\tau=1$, and the panels
c) and f) the vertical magnetic field at $\tau=1$. While the snapshot with 
$\langle\vert B_z\vert\rangle(\tau=1) = 8.4$~G shows magnetic field organized on scales close 
to the grid spacing of the simulation, the snapshot with 
$\langle\vert B_z\vert\rangle(\tau=1) = 86$~G shows magnetic field organized more on the
scale of granular downflows with a mostly sheet-like appearance. Several downflow lanes show
sheets with opposite polarity nearby. Panel d) shows also several brightness enhancements 
associated with strong field on scales of $100$~km and less. We provide also 2 animations of
Figure \ref{fig:IVzBz_kin_sat} in the online material (one for the kinematic and one for the saturated
phase). These animation show the same quantities as presented in Figure \ref{fig:IVzBz_kin_sat}.

Figure \ref{fig:pwr_pdf_sat}a) shows kinetic and magnetic energy spectra (at $\tau=1$), which were computed for
the $4$~km grid spacing case. As the solution is evolving from the kinematic growth phase to
the saturated regime, the peak of the magnetic energy spectrum is moving toward larger
scales. At the same time kinetic energy becomes suppressed by about a factor of $2$ on
scales smaller than $100$~km as a consequence of Lorentz-force feedback. We will discuss the saturation
process further in Section \ref{sec:transfer}. For the solution
reaching a vertical mean field strength of $80$~G in the photosphere, the magnetic energy is in
super-equipartition by about a factor of $2$ on scales smaller than $100$~km. 

For the case with $4$~km grid spacing presented here the e-folding time scale for magnetic energy in the photosphere
is about $50$ sec during the kinematic growth phase. The growth rate is strongly resolution dependent,
we find time scales of $120$, $350$, and $850$ sec for grid spacings of $8$, $16$, and $32$~km. This
leads on average to a resolution dependence of the kinematic growth rate $\gamma_K\sim \Delta x^{-1.36}$.
This resolution dependence is significantly steeper compared to simple estimates that yield for
$P_M \ll 1$ $\gamma_K\sim Re_M^{1/2}$, where $Re_M=v L/\eta$, $v$ and $L$ are typical velocity and length scales of 
the problem (for $P_M \gg 1$ $Re_M$ has to be replaced by $Re=v L/\nu$) . 
Since we do not have explicit viscosity and magnetic diffusivity we further assume that $Re_M$ is linked to the scale separation allowed for
by the numerical simulation.  Assuming a 5/3 Kolmogorov spectrum, we would expect $Re_M\sim \Delta x^{-4/3}$,
leading to $\gamma_K \sim \Delta x^{-2/3}$. The growth rate is more consistent with a $\gamma_K\sim Re_M$
dependence, which was also found by \citet{Pietarila-Graham:etal:2010:SSD}. Since we do not use here
any explicit numerical diffusivity a detailed interpretation is difficult. 
  
Figure \ref{fig:pwr_pdf_sat}b) shows the corresponding probability distribution functions (PDFs) for 
$\vert B\vert$ at $\tau=1$. The PDF has a peak at around $30$~G and a nearly exponential
drop for stronger field. For the snapshots with $\langle\vert B_z\vert\rangle>60$~G strong kG 
field concentrations cause a bulge for $\vert B\vert>1000$~G. In snapshots with
$\langle\vert B_z\vert\rangle=80$~G we find at $\tau=1$ field concentrations with more than
$2$~kG.

\begin{figure*}
  \centering 
  \resizebox{0.95\hsize}{!}{\includegraphics{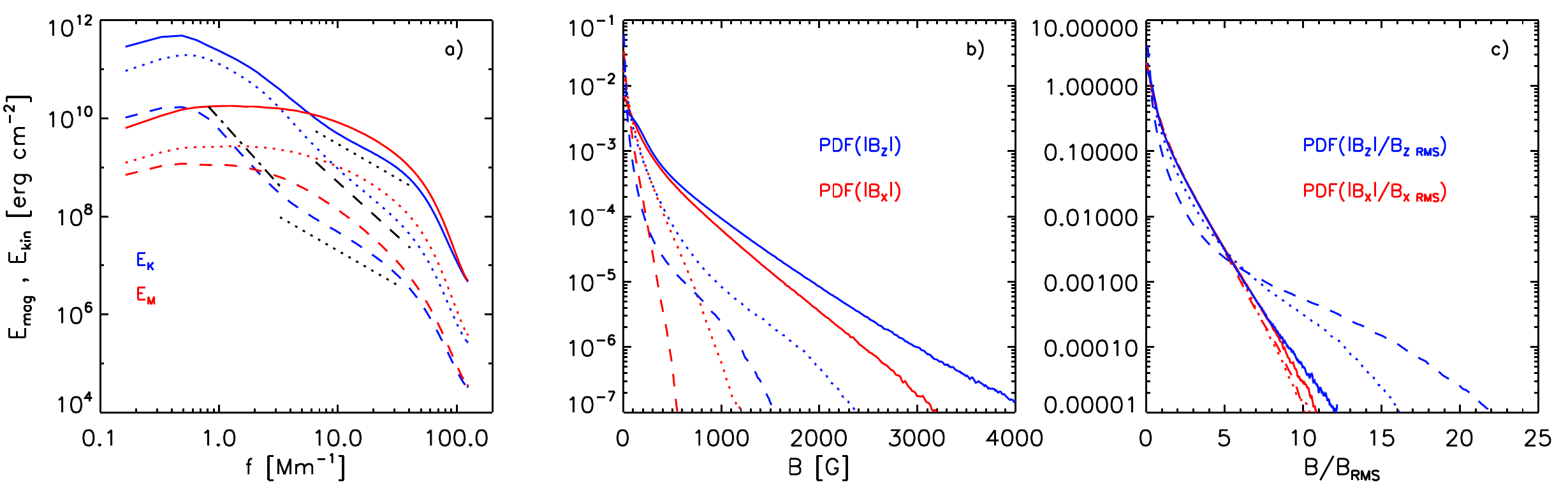}}
  \caption{Comparison of a) power spectra and b), c) probability
    distribution functions for $\vert B_x\vert$ and $\vert B_z\vert$ for
    3 different height levels in simulation O4a. Solid lines show quantities in $1$~Mm
    depth, dotted lines on the $\tau=1$ and dashed lines on the $\tau=0.01$
    levels. Panel c) compares normalized PDFs, which allow for a direct
    comparison of the shape regardless of field strength. In panel a) the
    black dotted, dashed and dashed-dotted lines indicate $-1.4$, $-2.2$ and $-2.7$
    power law slopes, respectively.
  }
  \label{fig:pwr_pdf_3lev}
\end{figure*}

\subsection{Resolution dependence}
Figure \ref{fig:pwr_pdf_res} compares simulations with 5 different grid spacings ranging from $32$ to $2$~km.
The simulation with $2$~km grid spacing was restarted from the saturated $4$~km simulation
and evolved for an additional $5$ minutes. Power spectra and PDFs were averaged over snapshots with
values of $\langle\vert B_z\vert\rangle(\tau=1)$  between $75$ and $85$~G. Panel a) shows kinetic
(blue) and magnetic (red) energy spectra for the simulations O32a-O2a. Increasing the resolution
leads to a convergence of the power spectra on the large scales while smaller scales are added. 
The simulations with $32$ and $16$~km grid spacing show excess power on large scales, since the same amount of magnetic energy
is distributed over less wave numbers. The simulations with $8$ to $2$~km grid spacing do not show a significant
difference indicating that a grid spacing of $8$~km or smaller is required to properly represent the energy
distribution on larger scales in the photosphere. The dotted line in panel a) indicates a Kolmogorov slope
of $-5/3$ as a rough reference.  Over the scale-range explored we don't see a clear
indication of a power law for the magnetic energy, there is some indication of power laws for the kinetic energy
(see also Figure \ref{fig:pwr_pdf_3lev}). Panel b) shows the ratio of magnetic to kinetic energy
as function of scale. For grid spacings smaller than $16$~km we find a super-equipartition regime on scales
smaller than $100$~km and see some indication that the ratio of magnetic to kinetic energy asymptotically
reaches a factor of about $2-2.5$. Panel c) shows the PDF for $\vert B_x\vert$ and $\vert B_z\vert$ at $\tau=1$. 
We do not see a systematic dependence on resolution, differences for stronger field are mostly realization noise. 
For field with less than $500$ G
strength the PDFs for $\vert B_x \vert$ and $\vert B_z \vert$ are essentially identical. Note that we show
here the PDFs for the absolute values of the field components since the simulations do not have any net magnetic
flux, leading to symmetric PDFs with respect to $B=0$.
 
Figure \ref{fig:energy_dist} shows normalized integrated magnetic energy spectra and distribution functions for
the simulations O32a-O2a (solid lines) . The quantities shown are defined as
\begin{eqnarray}
  S_k&=&\frac{\int_0^k E_M(k)dk}{\int_0^{k_{max}} E_M(k)dk}\\
  S_B&=&\frac{\int_0^B PDF(B) B^2dB}{\int_0^{B_{max}} PDF(B) B^2dB}
\end{eqnarray}
The quantity $S_k$ shows resolution dependence as expected from  Figure \ref{fig:pwr_pdf_res}a). In the highest
resolution case about $50\%$ of the magnetic energy in the photosphere on the $\tau=1$ level is found
on scales smaller than about $100$~km. Properly resolving the spectral magnetic energy distribution in the
photosphere requires grid spacings of $8$~km or smaller. In contrast to this the quantity $S_B$ shows
only little resolution dependence. In all cases $50\%$ of the magnetic energy is found in regions with
$\vert B \vert$ of less than $500$~G.  Kilo-Gauss field contributes about $10\%$ to the total magnetic energy.
For comparison we also show these quantities for the simulation
O16b (dotted) and O16bM (dashed). Both simulations have $25\%$ less unsigned flux in the photosphere.
The differences in $S_k$ are very minor. $S_B$ is shifted for the simulation O16b to the left toward weaker field. 
In contrast to that the simulation O16bM (larger domain) is very similar to the $80$~G cases and has an even larger 
contribution from kG field. We will discuss kG field concentrations further in Section \ref{sec:kG}.
      
\begin{figure*}
  \centering 
  \resizebox{0.95\hsize}{!}{\includegraphics{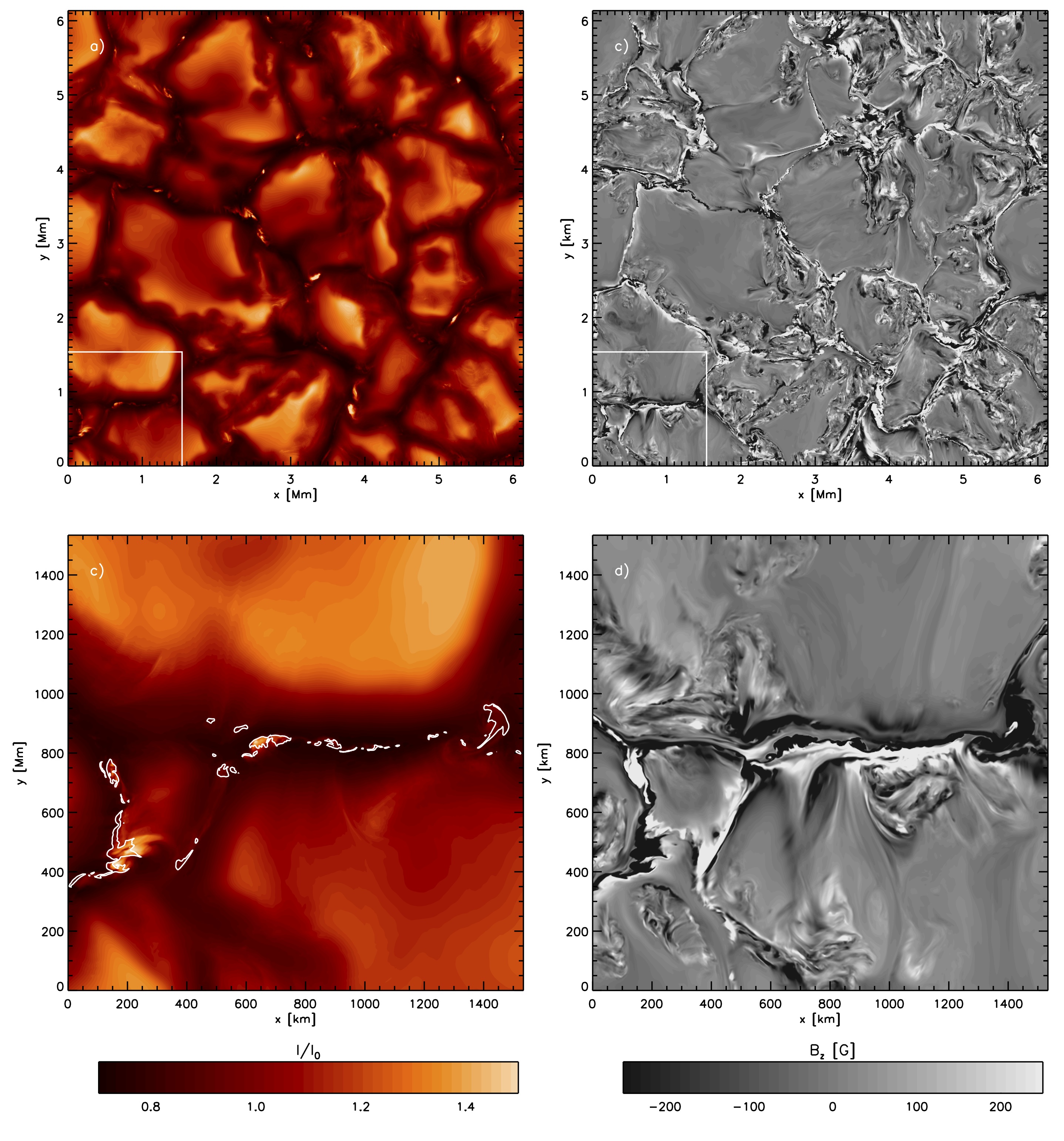}}
  \caption{Results from run O2a with a grid spacing of $2$~km. a) Intensity, b) $B_z(\tau=1)$, c) Intensity 
    patterns magnified for lower left corner of domain. Contour lines indicate regions with $\vert B\vert>1$~kG, 
    d)  $B_z(\tau=1)$ magnified for lower left corner.
  }
  \label{fig:IBz_2km}
\end{figure*}

\subsection{Height dependence}
\label{sec:height}
Figure \ref{fig:pwr_pdf_3lev} presents how power spectra and probability distribution functions for the
magnetic field strength depend on the vertical position in the simulation domain (based on run O4a). Here we focus on
three levels that are indicated by different line styles (solid: $-1$~Mm depth, dotted: $\tau=1$, 
dashed: $\tau=0.01$). Panel a) presents the kinetic and magnetic energy spectra for the three levels. They
show the same overall behavior with a super-equipartition regime toward small scales. While the super-equipartition
regime is reached at $\tau=1$ for scales smaller than $100$~km, it extends to $500$~km at $\tau=0.01$ since the 
kinetic energy drops more rapidly than magnetic energy above the photosphere (short density scale height). The 
super-equipartition regime extends also to moderately larger scales beneath the photosphere, since the overall
scale of convective motions increases with the increasing scale height, although the difference is small
between $\tau=1$ and $1$~Mm deeper. The black dotted, dashed and dashed-dotted lines indicate power law slopes 
of $-1.4$, $-2.2$ and $-2.7$, respectively. While we do not find a clear power law for $E_M$ at any height level there is some
indication of a power law for $E_k$ on scales smaller than downflow lanes (few $100$~km). At $1$~Mm depth
and $\tau=0.01$ we find slopes of about $-1.4$, while the $\tau=1$ level is with $-2.2$ substantially steeper.
Extrapolating the approximate slopes to smaller scales implies that the spectra of $E_k$ on the $\tau=1$ and $\tau=0.01$ levels 
will cross unless there is a change of slope toward smaller scales in either layer, which is more
likely. For all three layers we also find steeper slopes on scales larger than a few $100$~km. At $\tau=0.01$ we find
with $-2.7$ the steepest slope.

Comparing the PDFs for $\vert B_x\vert$ and $\vert B_z\vert$ (panel b) shows 
systematic differences in the overall shape between the distribution for vertical field at $\tau=1$ and 
$\tau=0.01$ and the rest. This difference is most obvious if we consider PDFs for the normalized magnetic field
components (panel c). Here the PDFs for $\vert B_x\vert$ at all three height levels and the PDF for $\vert B_z\vert$
in $1$~Mm depth are essentially identical, while the PDFs for $\vert B_z\vert$ at $\tau=1$ and $\tau=0.01$ show a 
much more extended tail toward stronger field. This is a strong indication for the presence of a distinct 
amplification process operating only on vertical field in the photosphere, while the distribution of $B_x$ in all 
three levels and $B_z$ beneath the photosphere is of mostly turbulent origin. While it is non-trivial to separate 
out the additional amplification process in the photosphere, we conjecture that it is related to a process along 
the lines of "convective intensification" \citep{Schuessler:1990:IAUS}, which is a combination of flux-expulsion,
back-reaction of magnetic field leading to partial evacuation, enhanced radiative cooling and related downflows.
These processes go beyond the idealized picture of "convective collapse" \citep{Spruit:1979:convcollapse}.

\begin{figure}
  \centering 
  \resizebox{0.85\hsize}{!}{\includegraphics{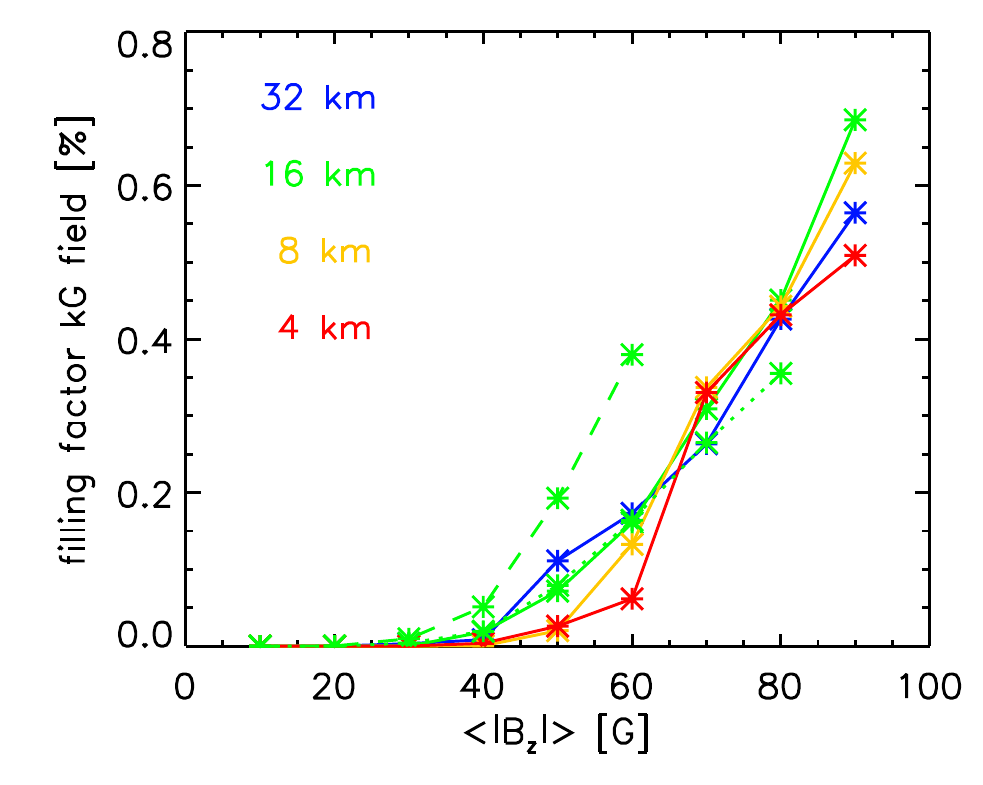}}
  \caption{Filling factor of kG field at $\tau=1$. Solid lines are computed from simulations
    O32a-O4a, the dotted (dashed) lines corresponds to O16b (O16bM). Field concentrations with more than 
    $1$~kG strength appear independent of resolution once $\langle\vert B_z\vert\rangle$ exceeds 
    about $30-40$~G. For $80$~G the filling factor reaches about $0.45\%$. This fraction is not 
    systematically dependent on resolution, but does increase with domain size (dashed line).
  }
  \label{fig:ff_kG}
\end{figure}

\begin{figure*}
  \centering 
 \resizebox{\hsize}{!}{\includegraphics{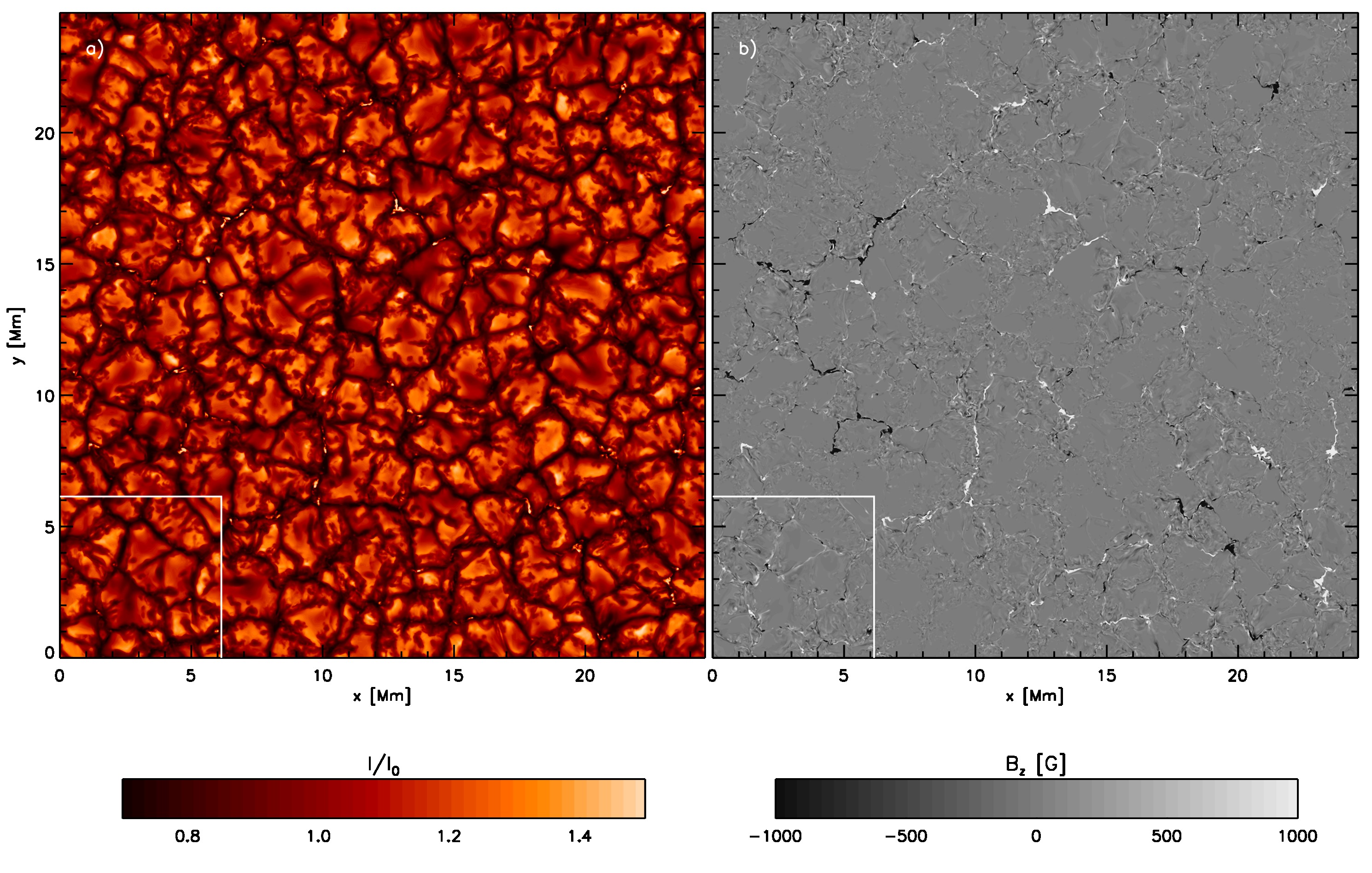}}
  \caption{a) Intensity and b) $B_z$ at $\tau=1$ taken from the simulation O16bM. The lower left corner 
    ($x,y<6.144$~Mm) shows $I$ and $B_z$ from the simulation O16b for comparison. The larger 
    domain shows indications of a network structure significantly larger than
    granulation that cannot form in the small domain. An animation of this figure displaying only O16bM is
    provided with the online material.
  }
  \label{fig:IBz_gran_meso}
\end{figure*}

\subsection{kG flux concentrations} 
\label{sec:kG}
Figure \ref{fig:IVzBz_kin_sat} shows the presence of several kG field concentrations in the photosphere
that lead to brightness enhancements in the downflow lanes. Here we analyze in more detail how
these flux concentrations depend on the overall field strength as well as domain size.

Figure \ref{fig:IBz_2km} shows examples of kG flux concentrations in the highest resolution simulation O2a. The panels a) 
and b) show $I$ and $B_z(\tau=1)$ for the full horizontal domain extent, while panels c) and d) show a magnification 
of the lower left corner of the domain. In panel c) contour lines highlight regions with $\vert B\vert>1$~kG. Many kG
field concentrations exist on scales smaller than $100$~km down to scales comparable to the grid resolution. Strong magnetic field 
is typically organized in sheets, often with alternating polarities. kG flux concentrations are small knots along these sheets in which
the field strength is increased temporarily due to dynamical effects. Some longer lived flux concentrations may be found in 
granular downflow vertices. We do not present here a detailed analysis of the temporal evolution of kG field concentrations, but refer
the interested reader to the animations of Figure \ref{fig:IVzBz_kin_sat} provided with the online material.  

Figure \ref{fig:ff_kG} presents how the filling factor of kG field at $\tau=1$ depends on the
vertical mean field strength, resolution as well as domain depth. To this end we computed for the simulations
O32a-O4a the filling factor of regions with $\vert B\vert > 1$~kG at $\tau=1$, while these 
simulations were evolving from the kinematic phase into the saturated phase. The data points presented in Figure
\ref{fig:ff_kG} result from binning snapshots in $\pm 5$~G intervals. For all 4 simulations 
we find regardless of the resolution that kG flux concentrations appear when the vertical mean field strength
exceeds about $30-40$~G at $\tau=1$. For $\langle\vert B_z\vert\rangle=80$~G around $0.45\%$ of the area 
is occupied by kG flux concentrations. While the results show some scatter due to realization noise in the simulation 
domain with small horizontal extent, there is no indication of a systematic resolution dependence of this result.

For comparison we also show the simulations O16b (dotted green) and O16bM (dashed green). While O16b
is comparable to O16a, for the same field strength O16bM shows about twice the filling factor. This difference
is related to the formation of a larger scale magnetic network structure we discuss further in Section \ref{sec:meso}.

\begin{figure*}
  \centering 
  \resizebox{0.95\hsize}{!}{\includegraphics{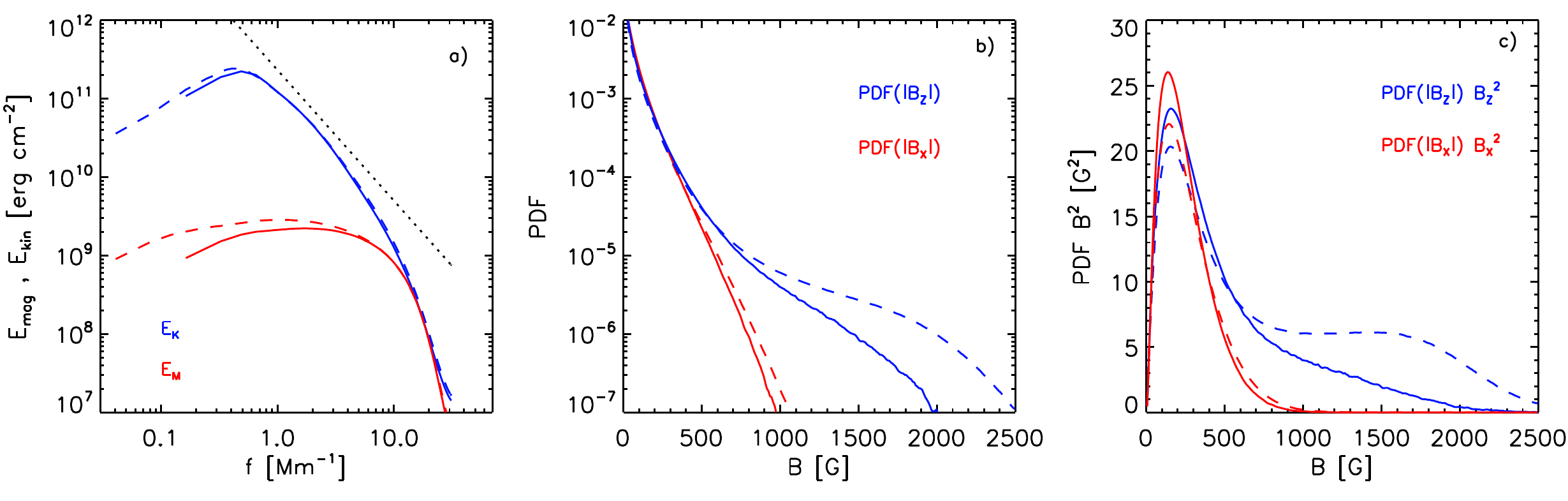}}
  \caption{a) Photospheric ($\tau=1$) power spectra for magnetic energy (red) 
    and kinetic energy (blue) in simulations O16b and O16bM (same resolution, but
    different domain size). Solid (dashed) lines refer to OS16b (OS16bM).
    b) Probability distribution functions for $\vert B_x\vert$ and
    $\vert B_z\vert$at $\tau=1$. Increasing the domain size increases the 
    magnetic power at larger scales and leads to a significantly higher fraction
    of field with more than $1$~kG field strength.
    c) Distribution of magnetic energy. In the larger domain the quantity
    $PDF(\vert B_z\vert)\, B_z^2$ shows a plateau toward $1700$~G.
  }
  \label{fig:pwr_pdf_gran_meso}
\end{figure*}

\subsection{From granular to meso-granular scales}
\label{sec:meso}
Figure \ref{fig:IBz_gran_meso} presents a comparison of snapshots from simulations O16b and O16bM.
Both simulations have a grid spacing of $16$~km and differ only in domain size. Presented are intensity and the $\tau=1$ magnetograms. 
In the larger domain magnetic field becomes organized on a scale larger then granulation. We find more pronounced kG flux 
concentrations that show up mostly as bright features in the intensity image. We do not find the spontaneous formation of larger 
pore-like field concentrations in O16bM. We provide also an animation of Figure \ref{fig:IBz_gran_meso} in the online material.
The animation shows only the simulation O16bM, but otherwise the same quantities as presented in Figure \ref{fig:IBz_gran_meso}.
Figure \ref{fig:pwr_pdf_gran_meso} compares the magnetic and kinetic energy spectra as well as probability
distribution function for the simulations O16b and O16bM.  We compare here time averages of snapshots with
values of $\langle\vert B_z\vert\rangle$ from $55$ to $65$~G.  In the larger domain the magnetic power spectrum extends 
toward larger scales, while the kinetic energy spectrum continues to fall off. We see an increase of magnetic power 
on scales larger than about $300$~km, while there is no significant change on smaller scales. 

The PDF  for $\vert B_x\vert$ remains mostly unchanged, while the PDF for $\vert B_z\vert$ shows a significant increase toward
kG fields in O16bM. We find that the filling factor of kG field is in O16bM with $0.4\%$ more than 
twice as large as in O16b ($0.16\%$). Computing the the distribution of energy from vertical magnetic field, $PDF(\vert B_z\vert)\, B_z^2$, 
leads to a plateau toward $1700$~G in O16bM that is not present in the smaller domain. We find the plateau only in the contribution from $B_z$.
In terms of the fraction of the total magnetic energy that is present in kG field at $\tau=1$ we find the values $9\%$ (O16b) and $23\%$ (O16bM).
In addition we studied also similar setups in larger domain ($98.304\times 98.304\times 18.432\,\mbox{Mm}^3$) at lower resolution and found 
that the trends indicated here (mostly flat magnetic energy spectrum on scales larger than granulation, increasing fraction of kG field) continue. 
In the simulation O32bSG we find a filling factor of $1.1\%$ for kG field, which contribute around $50\%$ to the magnetic energy at $\tau=1$.

The differences we see between the small and large domain arise from the presence of longer-lived, 
larger-scale convection flows present in the larger domain, which lead to the formation longer-lived flux 
concentrations. The trend of an increasing fraction of kG field with domain size indicates that perhaps even
a super-granular network structure could be maintained by a small-scale dynamo, provided the domain
is large enough. While the fraction of kG field increases with domain size, we did not find any indication 
for a secondary peak in the probability distribution functions of the magnetic field (including O32bSG).

\begin{figure*}[ht!]
  \centering 
  \resizebox{0.85\hsize}{!}{\includegraphics{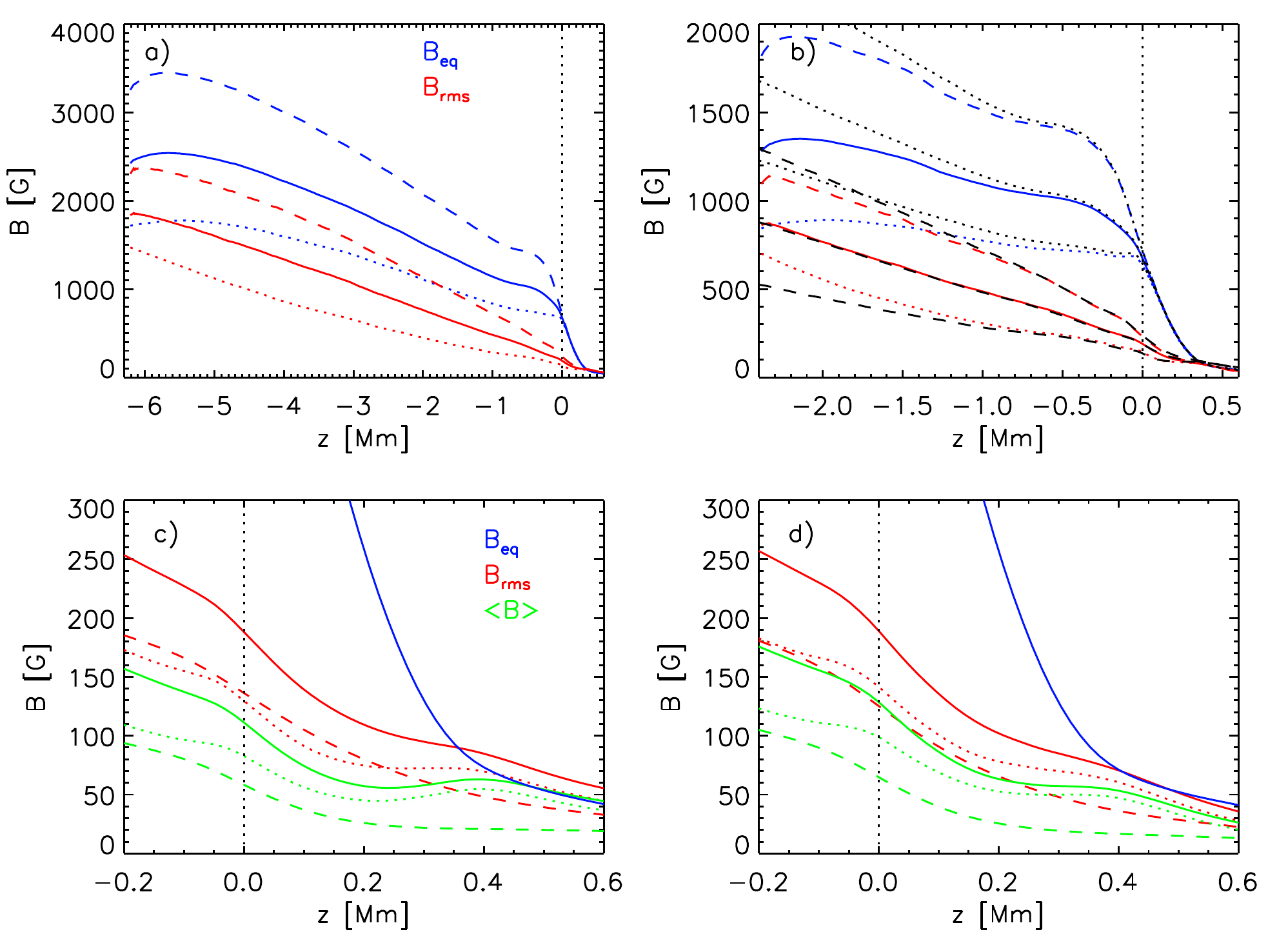}}
  \caption{Comparison of simulations in domains of different depth and
    horizontal extent. Both simulations use the boundary {\it OSb} and
    have a grid spacing of $16$~km (simulations O16b and O16bM).
    Panels a) and b) show $B_{\rm eq}$ and $B_{\rm RMS}$ as
    function of depth. Solid lines correspond to horizontal averages, 
    while dotted (dashed) lines are averages over 
    up- (down-) flow regions. Panel a) shows results for the simulation O16bM
    ($\sim 6.2$~Mm deep domain), panel b) the same quantities for simulation O16b
    ($\sim 2.3$~Mm deep domain). Both simulations are consistent with each other in terms
    of the RMS field strength.  Black (dotted) dashed lines 
    indicate in panel b) the profiles of $B_{\rm eq}$ and $B_{\rm RMS}$ from 
    panel a) for comparison.
    Panels c) and d) show the magnetic field structure at and above the
    photosphere. Red (green) lines indicate the RMS (mean) field strengths, while blue
    lines show the equipartition field strength. The meaning of line styles is different from
    panels a) and b): dashed (dotted) lines refer to the corresponding averages of vertical
    (horizontal) field components. 
  }
  \label{fig:BeqBrms}
\end{figure*}

\subsection{Subsurface field structure, role of boundary conditions}
\label{sec:subsurf}
\begin{figure*}
  \centering 
  \resizebox{0.85\hsize}{!}{\includegraphics{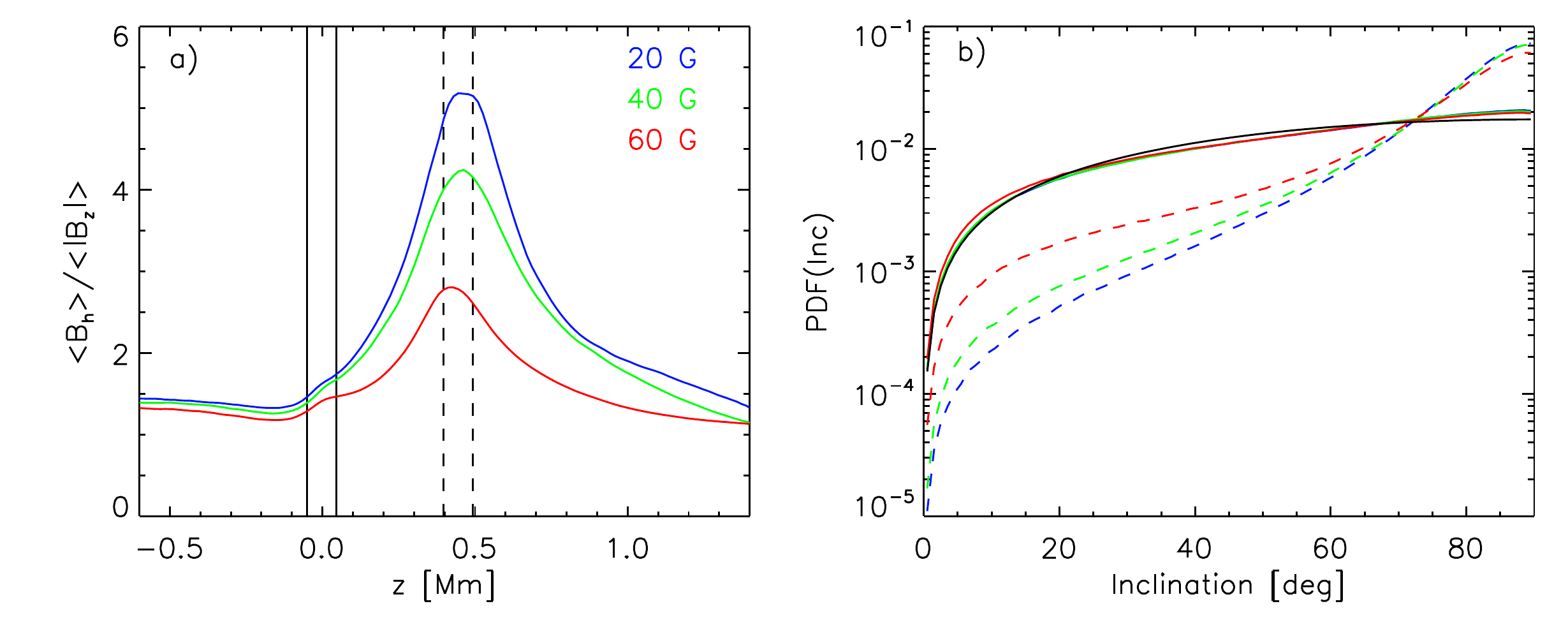}}
  \caption{a) Ratio of horizontal to vertical field strength as
    function of height. Different colors refer to simulations with the
    average vertical field strength at $\tau=1$ as indicated. The ratio
    of horizontal to vertical field has a maximum about $450$~km above
    $\tau=1$ and is strongly dependent on the overall field strength of
    the simulation and decreases with increasing field strength. b)
    Probability distribution functions for the field inclination with respect
    to the vertical. Solid lines refer to the deep photosphere around $\tau=1$,
    dashed lines to about $450$~km height as indicated in panel a). The black
    solid line indicates an isotropic distribution of field inclinations.
  }
  \label{fig:Inclination}
\end{figure*}

Figure \ref{fig:BeqBrms} presents for the simulations O16b and O16bM the vertical profiles of the equipartition field strength $B_{\rm eq}$ (blue) and $B_{\rm RMS}$ (red). The equipartition field strength $B_{\rm eq}=\sqrt{4\pi \varrho}\,v_{\rm RMS}$ is a measure for energy available in convective motions. 
Different line styles correspond to upflow regions (dotted),
downflow regions (dashed) and the averages over the whole domain (solid). In Panel b) black lines indicate the
profiles from Panel a) for better comparison ($B_{\rm RMS}$ is dashed, $B_{\rm eq}$ dotted).  
The simulations O16b and O16bM show a lot of similarity, in terms of the total $B_{\rm RMS}$ both simulations match 
each other in the part of the domain where they overlap. Differences are present when we compare $B_{\rm RMS}$ in
up and downflow regions in separation. Also $B_{\rm eq}$ is lower throughout most of the shallow domain, except for the 
near photospheric layers. The fact that the average magnetic properties in the shallow domain stay very
close to those in the deep domain is an indication that the bottom boundary condition {\it OSb} does perform fairly well
in "mimicking" a deep convection zone and leads to consistent results independent from the location of the
bottom boundary. The panels c) and d) give a more
detailed view of the magnetic field structure in and above the photosphere. For both simulations we find a secondary peak of the horizontal
mean field strength about $450$~km above $\tau=1$ (green dotted line). In the larger domain, panel c), 
this translates into a secondary peak of the total field
strength, while the RMS field strength continues to drop monotonically above the photosphere. We discuss the inclination 
of magnetic field above the photosphere in more detail in Figure \ref{fig:Inclination}.

\begin{figure*}[ht!]
  \centering 
  \resizebox{0.85\hsize}{!}{\includegraphics{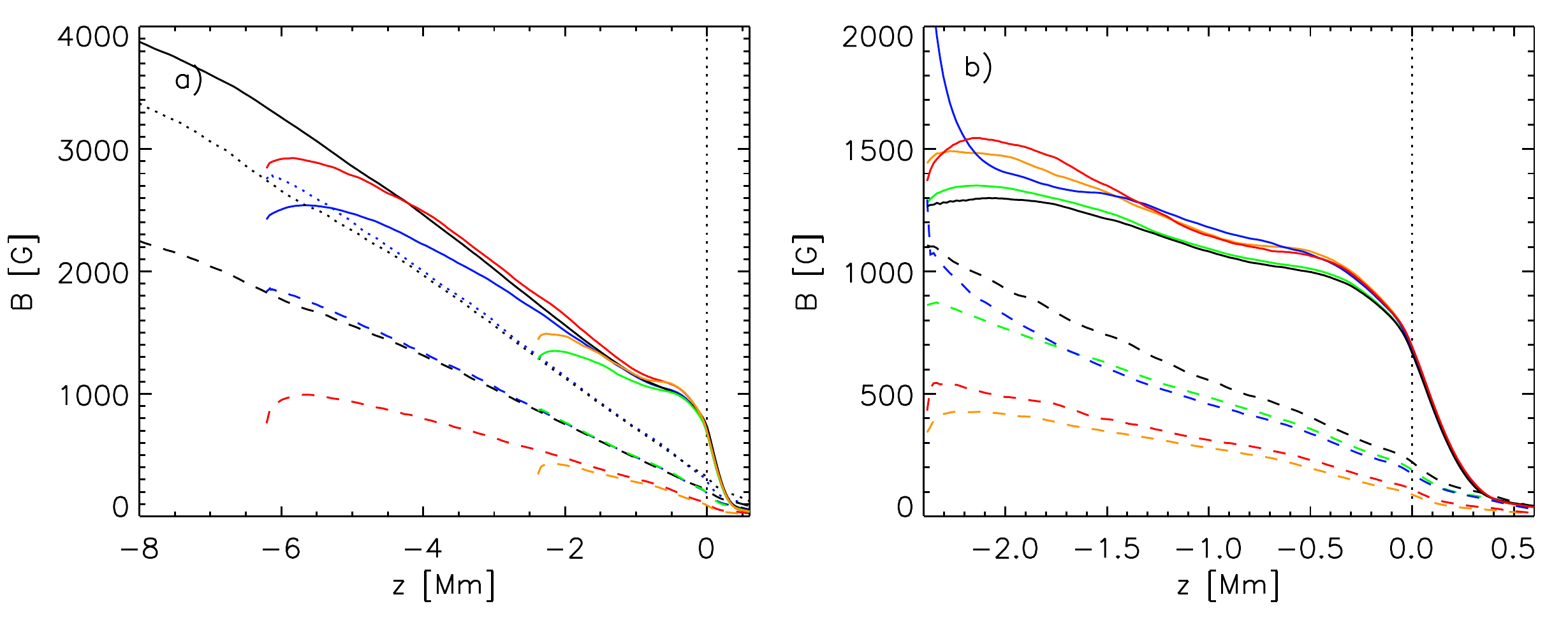}}
  \caption{Comparison of different bottom boundary conditions for a fixed grid spacing of $16$~km. 
    Solid (dashed) lines correspond to equipartition (RMS) field strength, the color indicates 
    simulations with different bottom boundary conditions. In panel a) we compare
    boundary {\it OSb}: O16bM (blue) and O16b (green) with boundary {\it OZ}: Z16M (red) and Z16 (orange).
    The black line show the results from O32bSG, for which the bottom boundary is in about $18$~Mm depth.
    In addition the dotted lines show scaled subsurface $B_{\rm RMS}$ profiles for O16bM and O32bSG to indicate a 
    solution we consider the upper limit ($B_{\rm RMS}$ increases with depth at the same rate as $B_{\rm eq}$). This
    solution corresponds to about $\langle\vert B_z\vert\rangle=85$~G at $\tau=1$ (based on O16bM).
    Panel b) presents additional experiments in the shallow domain.
    Here we present simulations with the boundaries  {\it OSa} (O16a, black), {\it OSb} (O16b, green), 
    {\it OZ} (Z16, orange), {\it OV} (V16, red) and {\it CH} (C16, blue).
  }
  \label{fig:BeqBrms_BND}
\end{figure*}

Figure \ref{fig:BeqBrms_BND} presents a comparison of simulations with different bottom boundary conditions. Panel a) compares
the boundaries {\it OSb} and {\it OZ} in the deep (runs O16bM and Z16M) and shallow domain (runs O16b and Z16). Changing from
{\it OSb} to {\it OZ} drops the field strength in the bulk of the convection zone by about a factor of $2$. The difference
between both boundary conditions does not depend on the domain depth within the range explored here. The boundary condition
{\it OZ} is very conservative in the sense that it assumes that the deeper convection zone is field free, which is unlikely
to be the case. But even with this assumption a still considerable amount of magnetic field is maintained within the
computational domain, although the mean vertical magnetic field at $\tau=1$ levels out at about $30$~G. This value is not
strongly dependent on resolution as long as a critical value is passed (i.e. the kinematic dynamo growth rate has to be
sufficiently large compared to the flux loss rate $\sim v_{z {\rm RMS}}/H_{\varrho}$).
We repeated this experiment with the resolutions from $32$ to $8$km.  While we find for $32$~km resolution only a vertical magnetic field 
strength of $20$~G at $\tau=1$, the simulations with $16$ and $8$~km grid spacing reach both values around $30$~G.

The simulations with the boundary condition {\it OSb} reach $>0.5 B_{\rm eq}$ in the deeper parts of the domain. These solutions are not
far from an upper bound for the field strength in which $B_{\rm RMS}$ and $B_{\rm eq}$ increase with depth at the same rate. To better illustrate 
this asymptotic limit we show also the results from O32bSG, which uses a $18$~Mm deep domain. The dotted lines indicate
$B_{\rm RMS}$  profiles for O16bM and O32bSG that are rescaled by a factor of $1.5$ to illustrate this asymptotic limit. Substantially stronger field 
would require a $B_{\rm RMS}$ increasing with depth faster than $B_{\rm eq}$ and exceeding equipartition in only a few Mm of depth.  This
asymptotic limit corresponds to a solution with $\langle\vert B_z\vert\rangle=85$~G,  $\langle B\rangle=160$~G, and $B_{\rm RMS}=275$~G  at $\tau=1$. 

\begin{figure*}[ht!]
  \centering 
  \resizebox{0.95\hsize}{!}{\includegraphics{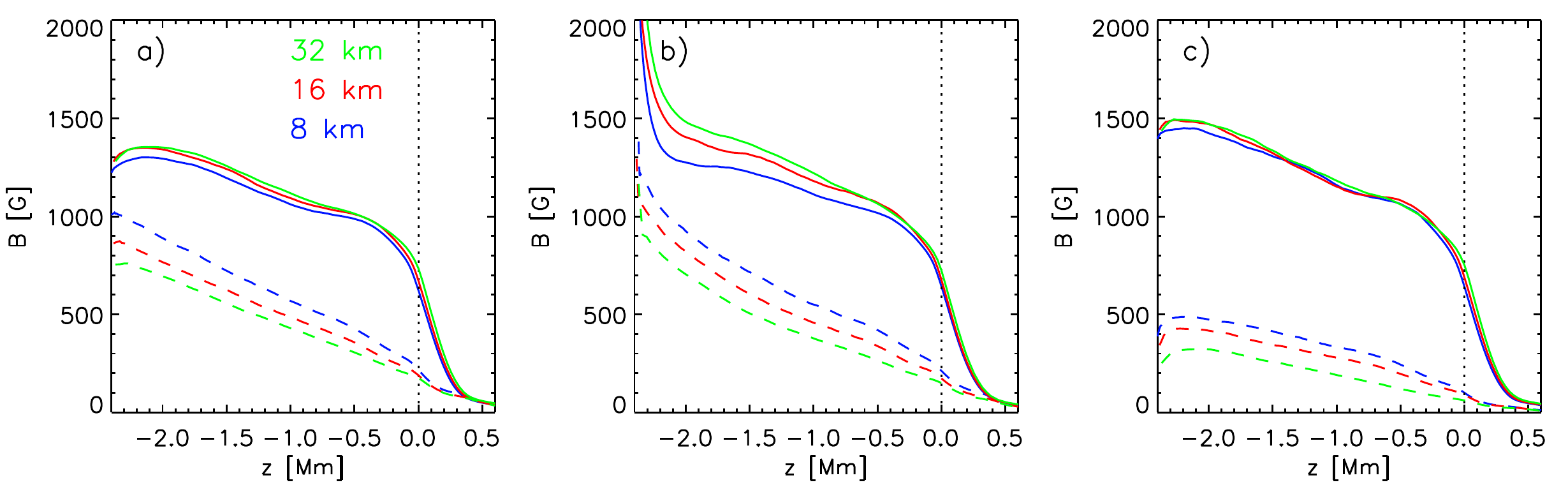}}
  \caption{Resolution dependence of $B_{\rm RMS}$ (dashed) and $B_{\rm eq}$ (solid) for simulations with the boundary
  a): {\it OSb}, runs O32b-O8b; b) {\it CH}, runs C32-C8; and c) {\it OZ}, runs Z32-Z8. 
}
\label{fig:BeqBrms_Res}
\end{figure*}

Figure \ref{fig:BeqBrms_BND}b) compares results of all 5 boundary conditions considered here for the shallow domain. We do not
find a significant difference between zero field in inflows {\it OZ} (orange) and vertical field everywhere at the bottom boundary {\it OV} (red). Due to the strong horizontal divergence in upflows, vertical magnetic field present at the 
bottom boundary condition becomes quickly expelled from upflows. Solutions with stronger magnetic field require the 
presence of horizontal field in upflow regions, which is less affected by horizontally divergent flows. The solutions with 
the boundary conditions {\it OSa,b} (black, green) are very similar to a solution computed with a closed bottom boundary condition {\it CH} (blue). The saturation field strength for the latter is fully determined by processes within the computational domain, while the former exchange magnetic field through the bottom boundary.

Figure \ref{fig:BeqBrms_Res} analyses further the resolution dependence of the saturation field strength for the simulations using the boundaries
{\it OSb}, {\it CH}, and {\it OZ}. In all three cases we find a similar trend of increasing $B_{\rm RMS}$ with resolution. While the saturation field
strength is not yet fully converged, it cannot grow much further in the simulations with the boundary conditions {\it OSb} and {\it CH} 
without creating a super-equipartition regime near the bottom of the domain.

\subsection{Subsurface Poynting flux and energy conversion rates}
\label{sec:energetics}
\begin{figure*}[ht!]
  \centering 
  \resizebox{0.95\hsize}{!}{\includegraphics{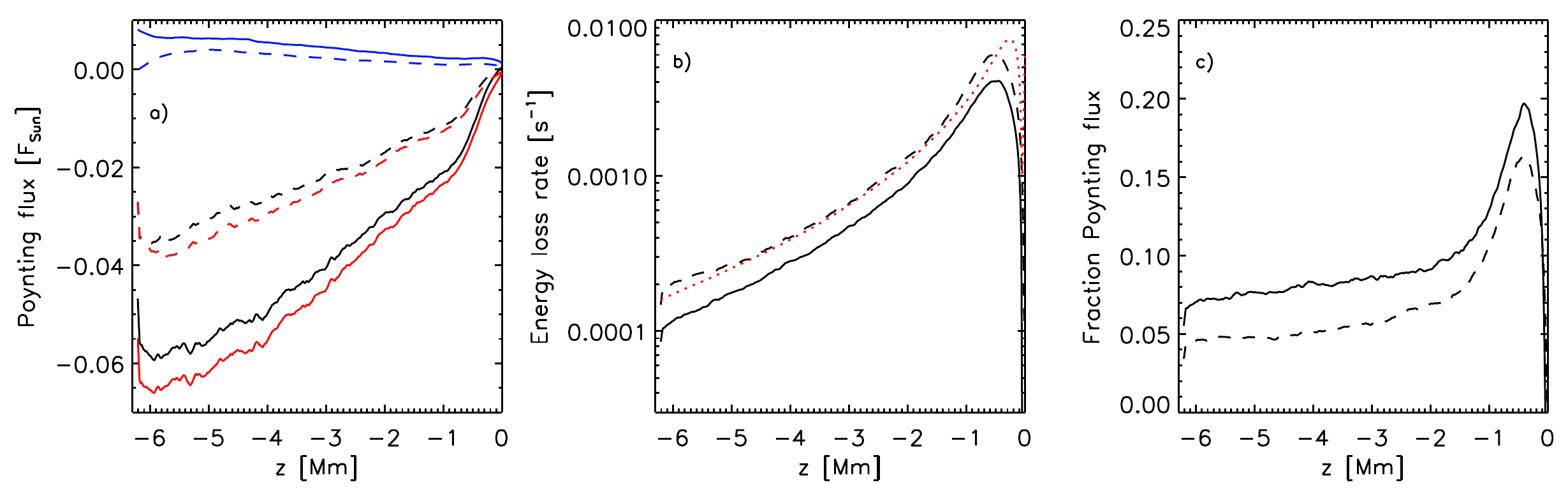}}
  \caption{Comparison of Poynting flux and associated time scales for the simulations O16bM (solid) and 
    Z16M (dashed). a) Black lines show the horizontally averaged Poynting flux, blue
    and red lines present the contributions from up and downflows. b) Energy loss
    rate due to Poynting flux ($\zeta(z)$, Eq. \ref{tau_poyn1}). The red dotted line indicates a convective overturning
    time scale $v_{z {\rm RMS}}/H_{\varrho}$. c) Fraction of energy transported by Poynting
    flux relative to energy converted by Lorentz force ($\varepsilon(z) $, Eq. \ref{tau_poyn2}).
  }
  \label{fig:Poynting}
\end{figure*}

Figure \ref{fig:Poynting}a) shows the Poynting flux for the
simulations O16bM and Z16bM. The flux is normalized by the solar photospheric energy flux of 
$F_{\odot}=6.3\cdot 10^{10}\,\mbox{erg}\,\mbox{cm}^{-2}\,\mbox{s}^{-1}$. Although we have magnetic energy entering the
domain in upflow regions in simulation O16bM (solid lines) using the boundary condition {\it OSb}, the magnetic energy leaving 
the domain in downflow regions over-compensates this contribution by more than a factor of $6$. In the simulation Z16M (dashed lines)
the Poynting flux is zero in upflow regions at the bottom boundary by construction (boundary condition {\it OZ}). However,
about $1-2$~Mm above the bottom boundary mixing between up and downflows provided enough field in upflow regions to have also
here an upward directed Poynting flux. The relative contributions from up and downflows in case Z16M are almost
identical with case O16bM if we stay $1-2$~Mm away from the bottom boundary condition. Panel b) shows the resulting magnetic energy loss rates for both simulations, which are defined as (with the Poynting flux $P(z)$)

\begin{equation}
  \zeta(z)=\frac{-P(z)}{\int_z^{z_{top}}E_M\,{\rm d}z}\;,\label{tau_poyn1}
\end{equation}
i.e. we compare the Poynting flux at a height $z$ to the total magnetic energy of the domain above $z$. For this analysis we can ignore
the Poynting flux at the top boundary, which is around $10^{-5} F_{\odot}$. In simulation Z16M with zero magnetic field in inflow regions 
we find an about $1.8$ times larger loss rate. The vertical profile of $\zeta$ agrees very well with a convective time scale
$v_{z {\rm RMS}}/H_{\varrho}$, indicated by a red dotted line. The kinematic growth phase of the dynamo is not affected by details of the
bottom boundary condition as long as the growth rate fullfils  $\gamma_K\gg v_{z {\rm RMS}}/H_{\varrho}$ (this condition is fulfilled well by
the higher resolution cases with a grid spacing of $8$~km or smaller, for the cases with $16$ and $32$~km grid spacing this condition is
fulfilled in the lower parts of the domain, but not in the photosphere). The bottom boundary matters when non-linear saturation
effects cause $\gamma(B) \rightarrow  v_{z {\rm RMS}}/H_{\varrho}$. The simulation Z16M presents a setup with the maximum possible energy loss
at the bottom boundary, since the time-scale of magnetic energy loss is the same as the time scale for mass exchange. In that sense
this setup presents a lower limit for an efficient dynamo with $\gamma_K\gg v_{z {\rm RMS}}/H_{\varrho}$ (a less efficient dynamo could have
of course an even lower saturation field strength). Stronger saturation field strengths require less leaky bottom boundary conditions, which require
the presence of a Poynting flux in upflow regions, like in O16bM.

Panel c) compares the energy lost by the Poynting flux to the energy converted via the Lorentz force in the domain above a height $z$:
\begin{equation}
  \varepsilon(z)=\frac{P(z)}{\int_z^{z_{top}}\vec{v}\cdot(\vec{j}\times\vec{B})\,{\rm d}z}\;.\label{tau_poyn2}
\end{equation}

\begin{figure*}
  \centering 
  \resizebox{0.95\hsize}{!}{\includegraphics{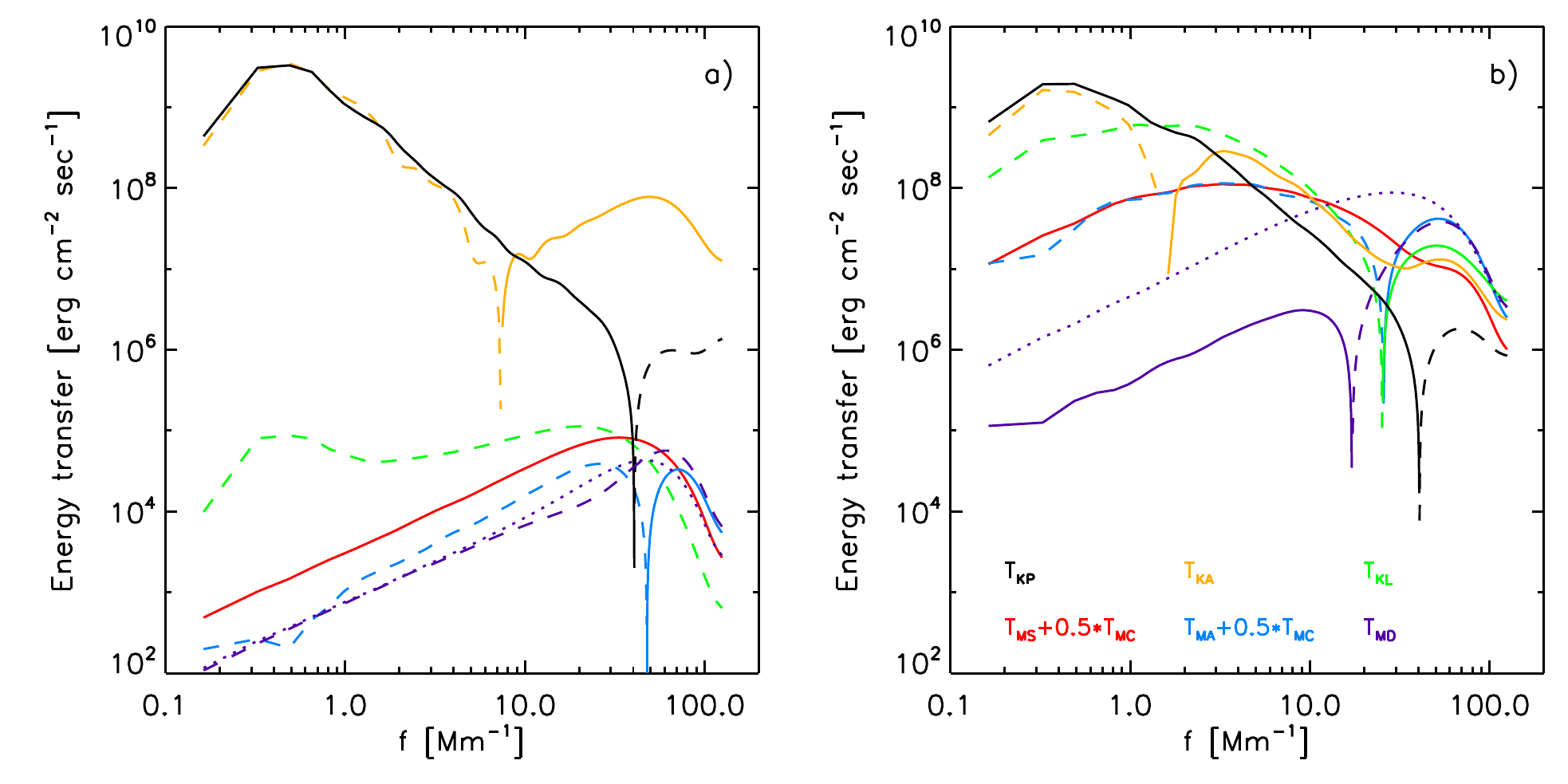}}
  \caption{Energy transfer functions for the simulation O16a with $4$~km grid spacing.
    Panel a) shows the kinematic growth phase, panel b) the saturated phase.
    The line color corresponds to contributions from pressure/buoyancy (black),
    advection of momentum (yellow), Lorentz force (green), stretching of magnetic field
    (red), and advection of magnetic field (blue). Contributions from the 
    magnetic compression term are split 50/50 among stretching and
    advection. Solid lines correspond to positive contributions, dashed lines
    to negative contributions. Contributions from numerical diffusivity in the 
    induction equation are indicated by the purple lines. Here the dotted line
    shows a Laplacian diffusivity with $\eta=5\cdot 10^{9}\,\mbox{cm}^2\mbox{s}^{-1}$
    for comparison.
  }
  \label{fig:Transfer_kin_sat}
\end{figure*}

\begin{figure*}[ht!]
  \centering 
  \resizebox{0.95\hsize}{!}{\includegraphics{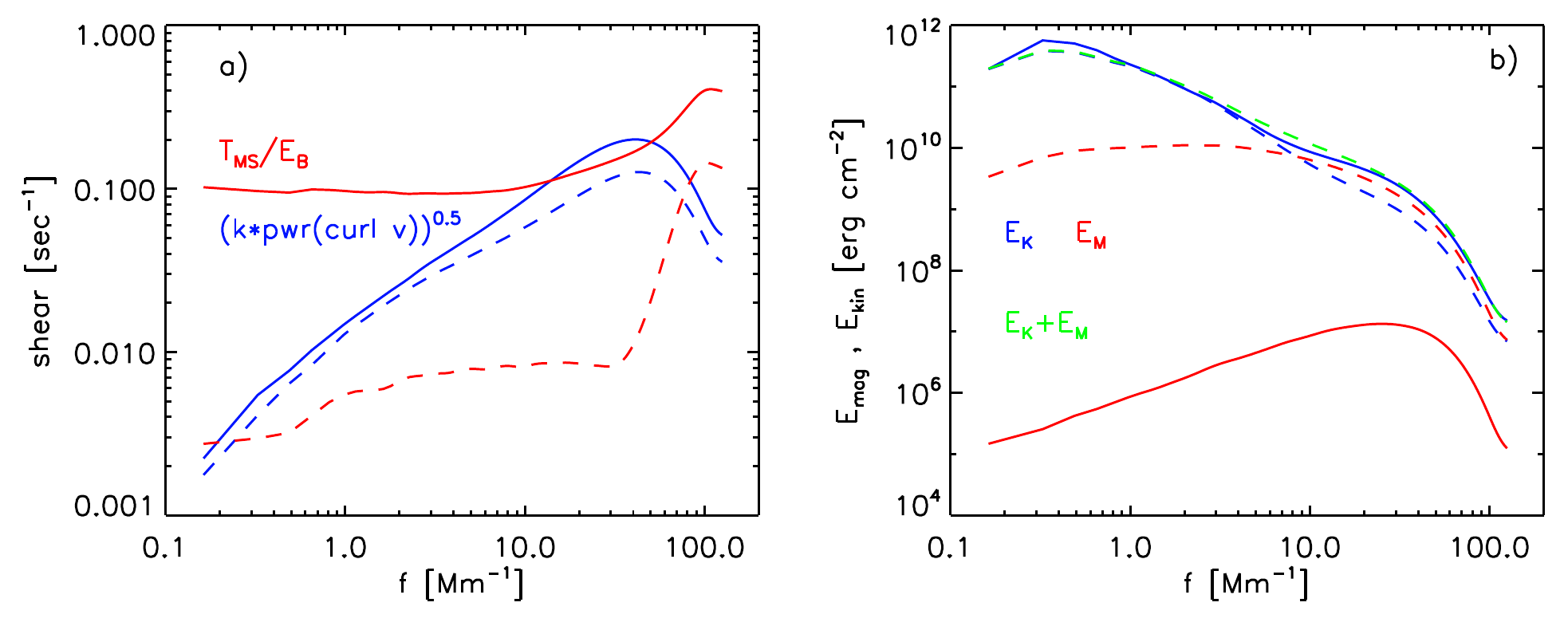}}
  \caption{Saturation of the dynamo comparing the kinematic growth phase with the
    saturated phase of O4a. Panel a) shows the vorticity spectrum $\sqrt{k\,P(\omega)}$ 
    (blue) and $T_{MS}/E_M$ (red). Solid (dashed) lines indicate the kinematic
    (saturated) phase. Panel b) shows the corresponding kinetic energy and magnetic energy 
    power spectra.}
  \label{fig:Saturation}
\end{figure*}

This fraction is lower in Z16M because of non-linear saturation effects in O16bM, which affect $\vec{v}\cdot(\vec{j}\times\vec{B})$ 
more strongly than $(\vec{v}\times\vec{B})\times\vec{B}$ (see also Figure \ref{fig:Saturation}).
While the simulation domain of O16bM contains almost $4$ times the magnetic energy
of Z16M, the average amount of energy converted from kinetic to magnetic energy is in both simulations comparable within a few 
$\%$, i.e. $85\,\mbox{erg}\,\mbox{cm}^{-3}\,\mbox{s}^{-1}$ (O16bM) and $80\,\mbox{erg}\,\mbox{cm}^{-3}\,\mbox{s}^{-1}$ (Z16M), 
which is about $50\%$ of the energy conversion by pressure/buoyancy forces in the domain, see also Section \ref{sec:transfer} for further detail.
Integrated over the depth of the domain this energy conversion rate equals to about $80\%$ of the energy flux through the domain. 
Note that we discuss here conversion rates between energy reservoirs and not true sinks of energy, since the energy is returned to 
internal energy through dissipation processes. The conversion rates can be comparable or even exceed the energy flux through the system. Most of the
energy converted from kinetic to magnetic is preferentially dissipated in downflow regions, while work against the Lorentz force reduces the kinetic
energy there. This changes the overall balance of convective energy transport by reducing the contribution from the kinetic energy flux. We find 
in a non-magnetic convection simulation in $6$~Mm depth a downward directed kinetic energy flux of about $-0.3\,F_{\odot}$, this value is 
reduced to $-0.2\,F_{\odot}$ in simulation O16bM. 

Recently \citet{Hotta:2014:Global_norot} presented small-scale dynamo simulations in a global setup covering the convection zone up to $7$~Mm
beneath the photosphere. Using a similar numerical approach, but a substantially lower grid spacing of $1100$~km horizontally and $375$~km vertically,
they were able to maintain a field with $0.15-0.25\,B_{\rm eq}$ throughout the convection zone. The maintenance of the field requires in their setup around
$5-10\,\mbox{erg}\,\mbox{cm}^{-3}\,\mbox{s}^{-1}$, which is consistent with our results considering the differences in the overall field strength reached
(their field near the top boundary falls short of our values by a factor of $4$, which is reflected in a more than a factor of $10$ lower energy conversion rate).
Integrated over the entire convection zone the energy conversion rate by a small-scale dynamo could account to as much as few
$L_{\odot}$ ($L_{\odot}=3.84\cdot 10^{33} \mbox{erg}\,\mbox{s}^{-1}$). Compared to that the energy extracted from large-scale mean flows in mean field dynamo
models \citep{Rempel:2006:dynamo} as well as 3D global dynamo simulations \citep{Nelson:2013:Wreaths&Cycles} is about 2 orders of magnitude smaller.

\subsection{Horizontal magnetic field above $\tau=1$}
In Figure \ref{fig:Inclination} we further analyze how the ratio of horizontal and vertical field
as well as distribution of inclination angles varies as function of height. In Panel a)
we present the quantity $\langle(B_x^2+B_y^2)^{0.5}\rangle/\langle\vert B_z\vert\rangle$ as
function of height for the simulation O16bM. Since this simulation was started from a weak
seed field we selected during the growth phase 3 snapshots with the field strength of
$\langle\vert B_z\vert\rangle=20$, $40$, and $60$~G at $\tau=1$. We find that independent 
from the field strength 
$\langle(B_x^2+B_y^2)^{0.5}\rangle/\langle\vert B_z\vert\rangle$ peaks about $450$~km above
$\tau=1$. The peak value reached drops monotonically with increasing field strength from about
$5$ at $20$~G to $2.75$ at $60$~G. The dashed vertical lines indicate regions for which we computed the 
PDFs for the inclination angle with respect to the vertical direction (panel b). Solid lines
refer to the PDFs around $\tau=1$, while dashed lines correspond to about $450$~km height.
For reference the black line indicates the distribution for an isotropic field. For all three
field strengths shown the PDFs are close to isotropic in the deep photosphere, but strongly
skewed toward horizontal field in $450$~km height.  The contribution from horizontal field
is strong enough to create a distinct peak in the field strength about $450$~km above $\tau=1$
as presented in Figure \ref{fig:BeqBrms}, panel c).  

\subsection{Transfer functions and saturation process}
\label{sec:transfer}

Figure \ref{fig:Transfer_kin_sat} presents energy transfer functions computed for the simulation O4a (4 km grid spacing).
We compare here the kinematic growth phase (panel a) with the saturated phase (panel b). 
The transfer functions are computed at a depth of about $800$~km beneath the photosphere.
They are averaged over a depth range of $160$~km. In addition we conducted a time average and applied
smoothing to the transfer functions in order
to suppress realization noise. Since the magnetic energy is varying rapidly during the 
kinematic growth phase we normalized the transfer functions that depend on the magnetic field
by the total magnetic energy in each time step and averaged the normalized transfer functions
(the kinematic growth is self-similar, i.e. only the amplitude and not the shape of the
transfer functions is changing). In Figure \ref{fig:Transfer_kin_sat}a) we scaled the corresponding 
transfer functions arbitrarily to show them on the same scale as the non-magnetic ones. 

Colors
refer to the transfer function defined in the Appendix. The energy transfers to the kinetic energy reservoir 
are $T_{KP}$ (black, energy transfer by pressure and buoyancy), $T_{KA}$ (orange, kinetic energy transfer by advection),
and $T_{KL}$ (green, kinetic energy transfer by the Lorentz force). The energy transfers to the magnetic energy
reservoir are $T_{MS}+0.5\,T_{MC}$ (red, energy transfer by stretching and compression), $T_{MA}+0.5\,T_{MC}$ 
(blue, energy transfer by advection and compression), and $T_{MD}$ (purple, energy transfer due to magnetic numerical
diffusivity). Solid (dashed) lines indicate positive (negative) contributions, the purple dotted lines indicate
the transfer of a Laplacian magnetic diffusivity with $\eta=5\cdot 10^{9}\,\mbox{cm}^2\mbox{s}^{-1}$ for
comparison (the simulations were only run with numerical diffusivity). We do not show the terms $T_{KD}$ for
better readability of the figures. We also split the term $T_{MC}$ $50/50$ among $T_{MA}$ and $T_{MS}$. The 
reason for this (apart from reducing the number of quantities shown in Figure \ref{fig:Transfer_kin_sat}) is that we can 
expand the underlying terms as:
\begin{eqnarray}
 -\vec{B}&\cdot&\left((\vec{v}\cdot\nabla)\vec{B}+\frac{\vec{B}}{2}\nabla\cdot\vec{v}\right)=
  -\nabla\cdot\left(\vec{v}\frac{B^2}{2}\right)\label{eq:tma}\\
  \vec{B}&\cdot&\left((\vec{B}\cdot\nabla)\vec{v}-\frac{\vec{B}}{2}\nabla\cdot\vec{v}\right)=
  \nabla\cdot\left(\vec{B}(\vec{v}\cdot\vec{B})-\vec{v}\frac{B^2}{2}\right)\nonumber\\
  &&-\vec{v}\cdot\nabla\cdot\left(\vec{B}\vec{B}-\frac{1}{2}\vec{I}\,B^2
      \right)\label{eq:tms}
\end{eqnarray}
i.e. the terms underlying $T_{MA}+0.5 T_{MC}$ (Eq. \ref{eq:tma}) can be identified with an advective energy
transport within the magnetic energy reservoir analogous to the terms underlying
$T_{KA}$ that refer to the turbulent momentum cascade. The terms underlying $T_{MS}+0.5 T_{MC}$ (Eq. \ref{eq:tms})
describe in part a transport within the magnetic energy reservoir (remaining non-advective
terms of Poynting flux) and in part the energy transfer with the kinetic energy reservoir (via Lorentz force).    

During the kinematic growth phase (panel a) $T_{MS}+0.5 T_{MC}$ peaks on a scale of about $25-30$~km, which
is about about $6-8\,\Delta x$, i.e. close to the smallest features that can be resolved with the given
grid spacing. With increasing resolution this scale is decreasing as it stays near $6-8\,\Delta x$. 
The corresponding Lorentz force related energy transfer $T_{KL}$ shows two peaks, one at around
$50$~km and one around $2$~Mm. The peak at around $50$~km is related to the magnetic tension force, while
the peak at $2$~Mm is caused by magnetic pressure. The dominant contribution to the energy exchange comes 
from the peak at small scales. On scales larger than $20$~km $T_{MS}+0.5 T_{MC}$ is partially opposed
by the transport term $T_{MA}+0.5 T_{MC}$ and a numerical diffusion term of similar amplitude. The remainder
leads to an exponential growth of magnetic energy with a e-folding time scale of about $50$~sec. On
scales smaller than $20$~km positive contributions from $T_{MS}+0.5 T_{MC}$ and $T_{MA}+0.5 T_{MC}$ are opposed
by numerical diffusivity. The contribution from numerical magnetic diffusivity (dashed purple line) is on scales
larger than $100$~km very similar to a Laplacian diffusivity with $\eta=5\cdot 10^{9}\,\mbox{cm}^2\mbox{s}^{-1}$
(dotted purple line), moderate differences exist on smaller scales. However, replacing our numerical diffusivity with
Laplacian diffusivity of  $\eta=5\cdot 10^{9}\,\mbox{cm}^2\mbox{s}^{-1}$ leads to an about $6$ times smaller kinematic
growth rate of the dynamo, which implies that it is non-trivial to estimate the effective numerical diffusivity by looking
at transfer functions or energy dissipation rates.

In the saturated phase (panel b) $T_{MS}+0.5 T_{MC}$ peaks on a scale of about $250-300$~km, about a factor of $10$
larger than during the kinematic growth phase. Similarly $T_{KL}$ peaks now at a scale of $500$~km, i.e. most of the
energy transfers from kinetic to magnetic energy happen on a scale comparable to downflow lanes. Unlike the kinematic 
growth phase these scales are independent of resolution and realized in all simulations presented here regardless of their 
resolution. On scales larger than $100$~km $T_{MS}+0.5 T_{MC}$ is in balance with $T_{MA}+0.5 T_{MC}$, contributions from
numerical diffusivity, $T_{MD}$, are about 2 orders of magnitude smaller. While $T_{MD}$ was close to Laplacian during the
kinematic growth phase, it differs substantially during the saturated phase. Contributions on scales larger than about
$70$~km are in amplitude about a factor of $10$ smaller than a Laplacian with $\eta=5\cdot 10^{9}\,\mbox{cm}^2\mbox{s}^{-1}$
(dotted purple line) and have the opposite sign. The latter is related to the cutoff we introduced in Eq. (\ref{phi_lim}) for
a setting of $h=2$. Integrated over all scales the positive contribution accounts to about $1.7\%$ of the total unsigned dissipation.
The sign change is not present for grid spacings of $8$~km and larger and can be avoided by using a setting
of $h=1$ in higher resolution cases (with no significant difference to the obtained results). On the smallest scales the
contribution from numerical diffusivity remains similar to a Laplacian diffusivity with $\eta=5\cdot 10^{9}\,\mbox{cm}^2\mbox{s}^{-1}$.  

\begin{figure*}[ht!]
  \centering 
  \resizebox{0.95\hsize}{!}{\includegraphics{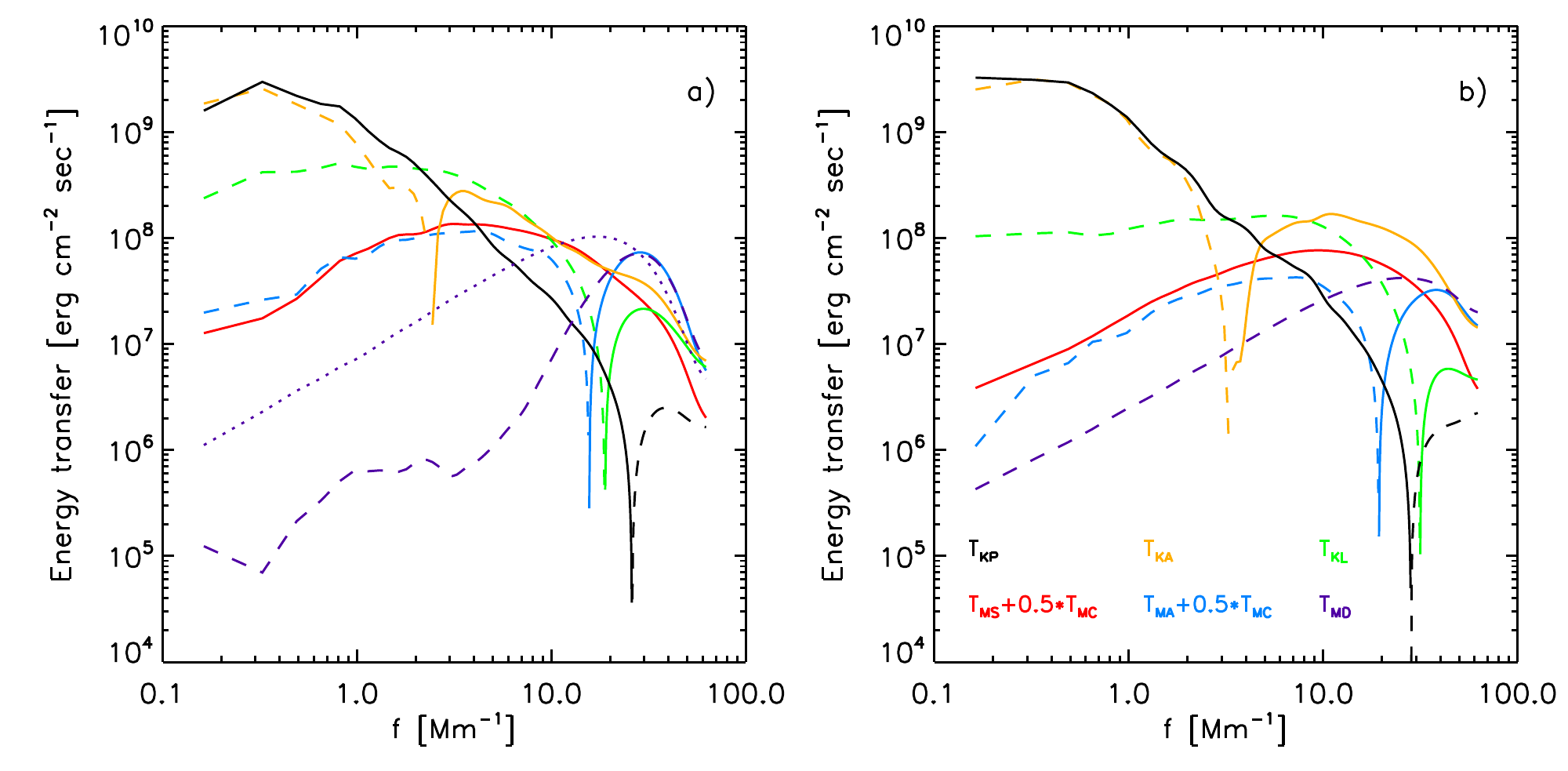}}
  \caption{Energy transfer functions for simulations C8 and C8$\eta$ with $8$~km grid spacing
    using the boundary {\it CH}. The meaning of the line styles is the same as
    in Figure \ref{fig:Transfer_kin_sat}.
    Panel a) shows a simulation computed with artificial magnetic 
    diffusivity, panel b) a simulation computed with a Laplacian
    diffusivity using a values of $\eta=10^{10}\,\mbox{cm}^2\mbox{s}^{-1}$. The 
    dotted purple line in panel a) indicates a Laplacian diffusivity with 
    $\eta=10^{10}\,\mbox{cm}^2\mbox{s}^{-1}$ for comparison.
  }
  \label{fig:Transfer_eta_hyp}
\end{figure*}

In the saturated phase the Lorentz force is a dominant contributor in the momentum equation and dominates energy transfers
on scales smaller than $500-1000$~km. While we find in the kinematic growth phase a balance between pressure/buoyancy driving,
$T_{KP}$, and the kinetic energy cascade, $T_{KA}$, down to scales of about $150$~km, this balance is only realized on scales
larger than $1000$~km in the saturated phase. From scales of $500$~km down to scales of $50$~km the term $T_{KA}$ balances
mostly $T_{KL}$ and as a consequence the amount of kinetic energy that is transported to small scales by $T_{KA}$ is significantly
reduced. Not all of the energy that is extracted by the Lorentz force is transferred into magnetic energy. On scales smaller 
than $40$~km $T_{KL}$ becomes the dominant source of kinetic energy. Overall $30\%$ of the energy extracted from kinetic energy on
large scales is returned to kinetic energy on small scales. Comparing transfer functions computed for simulations with different 
resolution, we find that this fraction increases with resolution. 

The total amount of energy that is dissipated numerically is within $10\%$ the same between kinematic and saturated phase.
In the saturated phase about $55\%$ of the energy dissipation happens through the magnetic channel. That we find about the same 
energy dissipation rates from magnetic and viscous dissipation is possibly related to an intrinsic numerical $P_M$ close to $1$.
\citet{Brandenburg:2011:SSD_low_Pm} found that this ratio depends on $P_M$ and that more energy is dissipated through the 
magnetic channel for low $P_M$.

Figure \ref{fig:Saturation}a) analyzes the saturation mechanism of the dynamo. To this end we compare the vorticity spectrum 
and the normalized transfer function $T_{MS}/E_M$ between kinematic growth phase and
saturated phase. We present these quantities averaged over the same depth range as the transfer functions in Figure \ref{fig:Transfer_kin_sat}.
We see only a moderate reduction of the overall shear by about $40\%$, mostly on smaller scales.
$T_{MS}/E_M$ drops by more than a factor of $10$, which indicates that most of the saturation happens though a 
misalignment of magnetic field and velocity shear. Panel b) shows the corresponding kinetic and magnetic energy spectra,
which are not very different from the photospheric spectra shown in Figure \ref{fig:pwr_pdf_sat}. In the saturated phase kinetic
energy is suppressed on scales smaller than $100$~km and magnetic energy is in super equipartition by about a factor of $1.7$.
The sum of kinetic and magnetic energy in the saturated phase is similar to the kinetic energy during the kinematic phase.

Figure \ref{fig:Transfer_eta_hyp} compares a solution computed with numerical diffusivity (panel a) to a solution computed
with a Laplacian magnetic diffusivity using $\eta=10^{10}\,\mbox{cm}^2\mbox{s}^{-1}$. These simulations are both
computed with a closed bottom boundary, i.e. the saturation level reached is determined by processes
within the computational domain and not sensitive to the magnetic bottom boundary condition. We present the
same quantities as in Figure \ref{fig:Transfer_kin_sat}. In panel a) the dotted purple line shows the transfer functions
of a Laplacian diffusivity with the same values as in panel b) for comparison. The solution computed with
a Laplacian diffusivity saturates at about half the field strength compared to the case with only numerical diffusivity.
The differences seen in the transfer functions for both cases are mostly a reflection of the differences in
overall field strength, i.e. we see the peak of $T_{MS}+0.5 T_{MC}$ shifted toward smaller scales as expected. 
The contribution from numerical diffusivity on large scales in panel a) is by about one order of magnitude smaller than the
contribution from the Laplacian diffusivity in panel b) (relative to the terms $T_{MS}+0.5 T_{MC}$. The value of 
$\eta=10^{10}\,\mbox{cm}^2\mbox{s}^{-1}$ is very close to the smallest value we can use for a grid spacing of $8$~km without 
excessive ringing and numerical instability on small scales. Achieving a regime similar to that shown in panel a) on large scales
using a Laplacian magnetic diffusivity would likely require about 10 times higher resolution, i.e. $10^4$ times more computing time. 

\section{Discussion}
\label{sec:discussion}
Since a significant fraction of the magnetic energy in the presented simulations is found on scales smaller than the resolution
of current instrumentation, a detailed comparison with observations is only possible through spectro-polarimetric
forward modeling. We defer such in-depth comparison to future publications, and limit the discussion here to a more
qualitative level.

\subsection{Field strength of quiet Sun}
We presented a series of simulations that lead to saturation field strength about a factor of $2-3$ higher than 
previously found by \citet{Voegler:Schuessler:2007}. We achieved that by a combination of a sufficiently high resolution 
in combination with an LES approach
(i.e. sufficient low numerical diffusivities) and different bottom boundary conditions. Higher resolution alone is
not sufficient, we found that the bottom boundary condition plays a crucial role. Using a conservative boundary assuming
zero magnetic field in inflow regions or only vertical magnetic field (similar to \citet{Voegler:Schuessler:2007}), we obtain 
a lower limit of about $\langle\vert B_z \vert\rangle=30$~G at $\tau=1$. The term "lower limit'' refers here to efficient 
dynamos, which have a growth rate $\gamma_K\gg v_{z {\rm RMS}}/H_{\varrho}$ during their kinematic phase. Less efficient dynamos
can have of course a lower saturation field strength. We derive an upper asymptotic limit of about $85$~G based on
the assumption that $B_{\rm RMS}$ cannot increase faster with depth than $B_{\rm eq}$. While our lower limit is possibly affected by 
the overall dynamo efficiency including potential magnetic Prandtl-number effects we do not account for, the upper limit
is only set by the available kinetic energy of the near surface convection zone. Similar values are also found in a setup with 
a closed bottom boundary, which, while less solar-like, provides a better posed dynamo problem. If we take our simulation
O16bM as basis for the  extrapolation, our "upper limit" of $\langle\vert B_z \vert\rangle=85$~G implies a value of 
$B_{\rm RMS}=275$~G, $\langle B \rangle=160$~G, and $\langle B_h \rangle=120$~G (all at $\tau=1$). These values are 
similar to those found by \citet{Danilovic:2010:zeeman_dynamo} through spectropolarimetric forward modeling
of rescaled dynamo models and comparison with observations ($\langle\vert B_z \vert\rangle=84$~G, $\langle B \rangle=170$).
We have repeated their analysis with non-grey versions of some of the simulations presented here (Danilovic $\&$ Rempel in prep.) and 
found that a model with $\langle\vert B_z \vert\rangle=60$~G agrees best with the data used in \citet{Danilovic:2010:zeeman_dynamo}.
The difference comes from the fact that due to non-linear feedback a rescaled weak-field solution is not identical to the fully non-linear 
strong field solutions we consider here. 
Inversion results by \citet{Orozco:Rubio:2012:QS_LP} ($\langle\vert B_z \vert\rangle=64$~G, $\langle B \rangle=220$) lead to 
similar values for $\langle\vert B_z \vert\rangle$, but significantly stronger field strength, which is due to an about a factor of $2$ 
stronger horizontal field ($198$~G) in their case.. 

\citet{Trujillo:etal:2004} and \citet{Shchukina:2011:hanle_dynamo} inferred from Hanle depolarization measurements values of 
$\langle B \rangle$ around $130$~G a few $100$~km above $\tau=1$. If we use our upper limit as reference, we find a field strength
in that height range around $80-90$~G, about a factor of $1.5$ less. If we use a model with $\langle\vert B_z \vert\rangle=60$~G 
(best fit to {\it Hinode} Zeeman data) the disagreement is almost a factor of $2$.

Overall these results indicate that the quiet Sun is magnetized near the upper limit we find, i.e. the observed field strength implies
that the subphotospheric layers have to be magnetized close to equipartition. The experiments with $B=0$ at the bottom boundary 
indicate that a small-scale dynamo restricted to the top $1-2$~Mm of the convection zone could only explain about $50\%$ of the 
observationally inferred field strength, i.e. the origin and strength of the quiet Sun magnetic field cannot be understood in separation
from the deeper layers of the convection zone.

\subsection{Scales beyond granulation}
\label{sect:disscuss_large_scale}
We presented one simulation on meso-granular scales using a domain $25$~Mm wide and $6$~Mm deep.  
Compared to the smaller domains in which no scales larger than granulation are allowed for, this simulation shows an organization 
of magnetic field on scales larger than granulation and also a stronger contribution from kG field. On a fundamental level this shows 
that the organization of magnetic field on a wide range of scales is not inconsistent with a small-scale dynamo, provided that the 
small-scale dynamo operates itself on a wide range of scales. The latter is a natural consequence of stratified convection in larger 
domains and does not require the contribution from a global dynamo, although that contribution becomes unavoidable once the 
domain size approaches scales on which rotation and shear become important. In that case a separation of contributions from a 
small-scale and large-scale dynamo is not trivial and not necessarily meaningful. 

\subsection{Coupling between the large- and small-scale dynamos}
\label{sect:discuss_LSD-SSD}
In our numerical experiments we focused entirely on setups with no imposed netflux, which leads to the question of how
the presence of a (cyclic) mean field produced by a large-scale dynamo might influence these results. We conducted an additional 
experiment (similar to O16bM) in which we imposed initially a vertical mean field of $30$~G. We found roughly a doubling of the 
photospheric magnetic energy, which is mostly due to the formation of a strong meso-granular network with $\vert B\vert >1$~kG, while 
the core of the probability distribution functions remains unchanged (the magnetic energy in regions with $\vert B\vert <500$~G 
is unchanged, in regions with $\vert B\vert<1$~kG we find only a $10\%$ increase). This result is expected since there is little
recirculation of mass in the top layers of the convection zone. The imposed netflux is expelled quickly into longer lived downflow
regions forming a magnetic network structure. The resulting inter-network regions are mostly void of netflux and have properties
similar to a simulation without any netflux. A similar weak dependence of the strength of inter-network field and the strength of the 
surrounding network field was also found by \citet{Lites:2011:hinode_ssd} in {\it Hinode} data. Overall this indicates that the strength
of mixed polarity field in the photosphere is only weakly influenceable by a vertical mean field. A much stronger modulation is possible 
through the properties of horizontal field in upflows regions, which is reflected in the strong dependence of our results on the details
of the bottom boundary condition. How much the strength of the quiet Sun is modulated by such a coupling depends ultimately
on how strongly the large-scale mean and small-scale mixed polarity field in the bulk of the convection zone vary throughout the 
cycle. Current global dynamo models do not have sufficient resolution to properly capture small-scale field and likely underestimate
its contribution. As discussed in Section \ref{sec:energetics} our estimates indicate that the energy converted by small-scale induction
effects exceeds the induction by large-scale mean flows (mostly differential rotation) by about two orders of magnitude. 

There are also possible feedbacks from the small-scale on the large-scale dynamo as well as convective dynamics. The presence of
small-scale field suppresses turbulent motions and reduces the kinetic energy flux. The amount of energy taken out of convective
motions through the Lorentz-force is substantial, if we equate this energy transfer with a viscous energy transfer we would require
an effective viscosity of a few times $10^{11}\mbox{cm}^2\mbox{s}^{-1}$ to mimic this effect. Recently \citet{Fan:Fang:2014:dynamo}
showed that the presence of magnetic field in the convection zone can be crucial for maintaining a solar-like differential rotation and
that the contribution of the magnetic field can be approximated to some degree by an enhanced effective viscosity of the flow. To which
degree small- and large-scale field components contribute to this effects requires further investigation.

\subsection{Distribution functions, kG field concentrations}
\label{sec:discuss_PDF}
We find that probability distribution functions are very robust, i.e. they barely depend on numerical resolution (we varied
the grid spacing by a factor of $16$!). The shape of PDFs is mostly determined by the average field strength and domain size.
For a given average field strength we find more strong field in a larger domain. Comparing photospheres with 
$\langle\vert B_z \vert\rangle=80$~G we find that $50\%$ of the energy comes from field with less than $500$~G, kG field
concentrations contribute about $16\%$ to the total energy. The latter drops to $9\%$ for a $60$~G case and increases to
$23\%$ for a $60$~G case in a larger domain. In our simulations the filling factor of kG field concentrations is strongly
field strength dependent.  More than $\langle\vert B_z \vert\rangle\sim 30-40$~G is required at $\tau=1$ before they form and
the filling factor increases steeply as the field strength increases beyond that threshold. 
Comparing the shape of normalized PDFs for vertical and
horizontal field components we find that the PDFs for vertical field in the photosphere deviate substantially from those of
horizontal field as well as vertical field beneath the photosphere (see Figure \ref{fig:pwr_pdf_3lev}). This is a strong hint for
the presence of a convective intensification mechanism \citep{Schuessler:1990:IAUS} that is restricted to the photosphere and
mostly affects vertical field. While we find strong field at $\tau=1$ reaching up $2.5$~kG, this field component is not organized
in form of flux tubes, nor does it have a preferred scale around $100$~km. Strong magnetic field  is typically organized in sheets, 
often with alternating polarities. kG flux concentrations are small knots along these sheets in which the field strength is increased
temporarily due to dynamical effects. We find kG flux concentrations down to the smallest scales we can resolve. The kG field present
in our simulations does not produce a distinct feature in the PDFs like a secondary peak around kG field strength found in many
observations \citep{Ishikawa:2009:cmp_QS_plage,Lites:2011:hinode_ssd,Orozco:Rubio:2012:QS_LP}. These observation also indicate that 
the second peak is possibly caused by contributions from network field and may not be present for inter-network field alone.  

\subsection{Power spectra}
\label{sec:discuss_power}
In our highest resolution simulation (grid spacing of $2$~km) we find that about $50\%$ of the magnetic energy in the deep
photosphere is found on scales smaller than $100$~km. Therefore properly reproducing the spectral energy distribution 
requires the highest possible resolution. Performing simulations with lower resolution will artificially move magnetic energy
toward larger scales in spectral space, for example we find $50\%$ on scales smaller than $300$~km for a grid spacing
of $32$~km. On scales smaller than $100$~km magnetic energy is in super-equipartition buy about a factor of $2$. A similar
feature has been found in several small-scale dynamo simulations and LES models of MHD turbulence 
\citep[see, e.g., review by][]{Brandenburg:2012SSRv}. We further find that the sum of kinetic and magnetic energy power spectrum
in the saturated state is similar to the kinetic energy power spectrum of a pure HD simulation, i.e. kinetic energy is suppressed 
on scales smaller than $100$~km and that gap is filled with magnetic energy. Even in the highest resolution case it is difficult
to identify power laws, in particular for magnetic energy. We see some indication for power laws in the kinetic energy.  Steeper
slopes (as steep as $-2.7$) are typically found for scales larger than a few $100$~km (width of downflow lanes). On smaller scales 
the slope is height dependent, we find $-1.4$ at $\tau=0.01$ and $1$~Mm beneath $\tau=1$, while we see a steeper 
$-2.2$ slope at $\tau=1$ in between these layers. This likely indicates that the slope at $\tau=1$ will change when 
approaching smaller scale, otherwise the kinetic energy at $\tau=1$ would drop below that at $\tau=0.01$. Recently
\citet{Katsukawa:2012:power} derived power spectra for kinetic and magnetic energy from {\it Hinode} data. They found
in the frequency range from $1.5$ to $3.5\,\mbox{Mm}^{-1}$ kinetic energy spectra with slopes around $-3.3$ to $-3.6$, 
while the slope of the magnetic energy spectrum is less steep with a slope around $-1.4$. While a detailed comparison
of these slopes is likely difficult without properly accounting for resolution and noise effects, we see in our simulations at 
least some indication for the substantially different slopes for kinetic and magnetic energy. In the frequency range
 $1.5$ to $3.5\,\mbox{Mm}^{-1}$  we find at $\tau=1$ and $0.01$ steep kinetic energy spectra with slopes as steep as
$-2.7$, while magnetic energy spectra are flat with slopes of less than $-1$. Based on these simulations we caution not to
extrapolate these slopes as they still change significantly toward smaller scales.

We see no indication that kG magnetic field present in our simulations would create a secondary peak in the magnetic power spectrum
around $100$~km as suggested by \citet{Stenflo:2012:scaling}.      

\subsection{Field inclination}
We find in the deep photosphere a close to isotropic magnetic field distribution, while higher layers are dominated by horizontal field. 
The ratio of horizontal to vertical field peaks about $450$~km above $\tau=1$. The exact value of this ratio is strongly field strength
dependent. While we can find ratios as high as $5$ for a solution with $\langle\vert B_z \vert\rangle=20$~G,
the ratio drops to less than $3$ for field strength values that are most compatible with observations. Over the height range where for
example {\it Hinode} observations are taken the ratio is more close to $2$. These values are consistent with those reported by 
\citet{Schuessler:Voegler:2008:bhorz} when we take into account that they considered a dynamo model reaching only 
$\langle\vert B_z \vert\rangle=25$~G, and that they used in their estimates the horizontal RMS field strength which is up to a
factor of $2$ stronger than the mean horizontal field we use. On the observational side \cite{Lites:etal:2008} found a ratio of $5$, while 
recently \citet{Orozco:Rubio:2012:QS_LP} deduced a lower value of $3.1$. In contrast to that other investigations such as \citet{MartinezGonzales:2008:iso,AsensioRamos:2009:iso} find a mostly isotropic field.  Recently \citet{Stenflo:2013:inc} showed that
the angular distribution of magnetic field in the quiet Sun varies with height. While the deep photosphere is more vertical, the upper photosphere
tends to be more horizontal. At least on a qualitative level we find a similar result, the deep photosphere is close to isotropic and higher
layers are more horizontally inclined, with the most horizontal distribution found $450$~km above $\tau=1$.

\subsection{Spectral energy transfers}
We computed for our models spectral energy transfer functions in order to analyze in detail the operation of the dynamo during kinematic
and saturated phase.
During the kinematic growth phase most energy transfers happen on scales $\sim 6-8\,\Delta x$, i.e. this scale is resolution
dependent. Overall the transfer functions for the kinematic growth phase are similar to those presented by 
\citet{Pietarila-Graham:etal:2010:SSD}. In the saturated phase energy transfers happen on a scale of  $250-500$~km, i.e. the 
scale of downflow lanes. This scale is found independent of the adopted numerical resolution. Contributions from numerical diffusivity 
on scales larger than $10\,\Delta x$~km are negligible and a balance between stretching and non-linear transport terms is achieved, i.e. 
the regime expected for $R_M\gg 1$ (which is the main reason for using an LES approach in the first place). In the saturated phase the
Lorentz force becomes the major player in the momentum equation on scales smaller than $1$~Mm.
Comparing models using physical and numerical magnetic diffusivity we do not find a significant difference in the transfer functions, but
differences exist in the kinematic growth rate (about $6$ times larger with numerical diffusivity) and saturation field strength 
(about $2$ times stronger with numerical diffusivity). 
Ultimately the validity of a LES approach for a small-scale dynamo problem has to be tested 
against high resolution DNS models. While the latter is becoming feasible for simplified setups, it will remain a substantial challenge for realistic 
solar-like MHD simulations as discussed here. 

\section{Concluding remarks}
\label{sect:conclusion}
We presented a series of MHD simulations of the solar photosphere with the aim of understanding better the origin of
small-scale mixed polarity magnetic field in the solar photosphere. Such field is ubiquitous on the solar surface and provides the 
dominant contribution to the magnetic energy in the quiet Sun photosphere. Previous simulations of small-scale dynamos fell short by a factor of
$2-3$ in terms of the field strength required to explain the observed level of Zeeman polarisation\citep{Danilovic:2010:zeeman_dynamo}.
In our study we showed that increasing resolution and reducing numerical diffusivities alone is not sufficient to reach the observationally
inferred field strength. The saturated solutions remain strongly dependent on the (open) bottom boundary condition typically used in photospheric
MHD simulations (note: the kinematic growth phase is not very sensitive to the choice of the bottom boundary condition). Physically this boundary 
condition dependence implies a strong coupling between the photosphere and the deeper
convection zone, i.e. the magnetism of the photosphere cannot be understood in separation from the rest of the convection zone.
Solutions that agree with observational constraints on the field strength in the photosphere imply a subsurface magnetic field with an
energy density comparable to the kinetic energy density. By that measure the small-scale magnetic field is far from being a weak field.
The energy conversion through the associated Maxwell-stresses accounts to about a solar luminosity when integrated over the top $10$~Mm
of the convection zone, which is about $50\%$ of the energy conversion by pressure/buoyancy forces. The Lorentz-force feedback on the flow 
leads to a significant reduction of kinetic energy flux as well as kinetic energy on small scales. Such feedback has potentially dynamical consequences
for the convective dynamics of the convection zone including the maintenance of mean flows and operation of a large-scale dynamo
\citep{Fan:Fang:2014:dynamo}.   

We further studied how the strength and distribution of magnetic field in the photosphere is influence by numerical resolution. Since about
$50\%$ of the magnetic energy resides in the deep photosphere on scale smaller than $100$~km, a sufficiently high numerical resolution
(ideally a grid spacing smaller than $8$~km) is required for properly capturing the spectral energy distribution. Probability distribution
functions of the magnetic field strength are on the other hand insensitive to numerical resolution in the explored range (grid spacings from
$32$ down to $2$~km) and can be considered converged. Our models are all based on the strong assumption that an LES approach
(using only numerical diffusivities) with an intrinsic $P_M$ close to unity is a proper way to deal with the regime found on the Sun. Within 
this approach results are very consistent and robust, which implies that they are either correct or systematically wrong. A detailed comparison 
with observations through forward modeling, which is work in progress, should tell which is the case.  

\acknowledgements 
The National Center for Atmospheric Research (NCAR) is sponsored by the National Science Foundation. The author thanks M. Sch{\"u}ssler,
A. DeWijn and Kyle Augustson for helpful discussions and comments on the manuscript. We would like to acknowledge high-performance computing support from Yellowstone 
( http://n2t.net/ark:/85065/d7wd3xhc) provided by NCAR's Computational and Information Systems Laboratory, sponsored by the National 
Science Foundation, under project NHAO0002; from the NASA High-End Computing (HEC) Program through the NASA Advanced
Supercomputing (NAS) Division at Ames Research Center under project s9025; and by the National 
Science Foundation and the University of Tennessee through the use of the Kraken computing  resource at the National Institute 
for Computational Sciences (http://www.nics.tennessee.edu) under grant AST100005.
This research has been partially supported through NASA contracts NNH09AK02I, NNH12CF68C and NASA grant NNX12AB35G. M. Rempel is grateful
to NAOJ for support of a Visiting Professorship during November 2011. Many fruitful discussions with the Japanese solar physics community during
that time sparked my interest in quiet Sun magnetism.

\bibliographystyle{natbib/apj}
\bibliography{natbib/papref_m}

\begin{thebibliography}{}
\expandafter\ifx\csname natexlab\endcsname\relax\def\natexlab#1{#1}\fi

\bibitem[{{Asensio Ramos}(2009)}]{AsensioRamos:2009:iso}
{Asensio Ramos}, A. 2009, \apj, 701, 1032

\bibitem[{{Bellot Rubio} \& {Orozco
  Su{\'a}rez}(2012)}]{Bellot:2012:pervasive_lin_pol}
{Bellot Rubio}, L.~R., \& {Orozco Su{\'a}rez}, D. 2012, \apj, 757, 19

\bibitem[{{Bercik} {et~al.}(2005){Bercik}, {Fisher}, {Johns-Krull}, \&
  {Abbett}}]{Bercik:2005:dynamo}
{Bercik}, D.~J., {Fisher}, G.~H., {Johns-Krull}, C.~M., \& {Abbett}, W.~P.
  2005, \apj, 631, 529

\bibitem[{{Brandenburg}(2011)}]{Brandenburg:2011:SSD_low_Pm}
{Brandenburg}, A. 2011, \apj, 741, 92

\bibitem[{{Brandenburg} {et~al.}(2012){Brandenburg}, {Sokoloff}, \&
  {Subramanian}}]{Brandenburg:2012SSRv}
{Brandenburg}, A., {Sokoloff}, D., \& {Subramanian}, K. 2012, \ssr, 169, 123

\bibitem[{{Buehler} {et~al.}(2013){Buehler}, {Lagg}, \&
  {Solanki}}]{Buehler:2013:cyc_dep}
{Buehler}, D., {Lagg}, A., \& {Solanki}, S.~K. 2013, \aap, 555, A33

\bibitem[{{Cattaneo}(1999)}]{Cattaneo:1999}
{Cattaneo}, F. 1999, \apj, 515, L39

\bibitem[{{Danilovic} {et~al.}(2010){Danilovic}, {Sch{\"u}ssler}, \&
  {Solanki}}]{Danilovic:2010:zeeman_dynamo}
{Danilovic}, S., {Sch{\"u}ssler}, M., \& {Solanki}, S.~K. 2010, \aap, 513, A1

\bibitem[{{de Wijn} {et~al.}(2009){de Wijn}, {Stenflo}, {Solanki}, \&
  {Tsuneta}}]{deWijn:2009:SSRv}
{de Wijn}, A.~G., {Stenflo}, J.~O., {Solanki}, S.~K., \& {Tsuneta}, S. 2009,
  \ssr, 144, 275

\bibitem[{{Dedner} {et~al.}(2002){Dedner}, {Kemm}, {Kr{\"o}ner}, {Munz},
  {Schnitzer}, \& {Wesenberg}}]{Dedner:etal:2002:divB}
{Dedner}, A., {Kemm}, F., {Kr{\"o}ner}, D., {et~al.} 2002, Journal of
  Computational Physics, 175, 645

\bibitem[{{Dom{\'{\i}}nguez Cerde{\~n}a} {et~al.}(2003){Dom{\'{\i}}nguez
  Cerde{\~n}a}, {Kneer}, \& {S{\'a}nchez Almeida}}]{Dominguez:2003:quiet_sun}
{Dom{\'{\i}}nguez Cerde{\~n}a}, I., {Kneer}, F., \& {S{\'a}nchez Almeida}, J.
  2003, \apjl, 582, L55

\bibitem[{{Dom{\'{\i}}nguez Cerde{\~n}a} {et~al.}(2006){Dom{\'{\i}}nguez
  Cerde{\~n}a}, {S{\'a}nchez Almeida}, \& {Kneer}}]{Dominguez:2006:quiet_sun}
{Dom{\'{\i}}nguez Cerde{\~n}a}, I., {S{\'a}nchez Almeida}, J., \& {Kneer}, F.
  2006, \apj, 646, 1421

\bibitem[{{Fan} \& {Fang}(2014)}]{Fan:Fang:2014:dynamo}
{Fan}, Y., \& {Fang}, F. 2014, ArXiv e-prints, arXiv:1405.3926

\bibitem[{{Hotta} {et~al.}(2014){Hotta}, {Rempel}, \&
  {Yokoyama}}]{Hotta:2014:Global_norot}
{Hotta}, H., {Rempel}, M., \& {Yokoyama}, T. 2014, \apj, 786, 24

\bibitem[{{Ishikawa} \& {Tsuneta}(2009)}]{Ishikawa:2009:cmp_QS_plage}
{Ishikawa}, R., \& {Tsuneta}, S. 2009, \aap, 495, 607

\bibitem[{{Iskakov} {et~al.}(2007){Iskakov}, {Schekochihin}, {Cowley},
  {McWilliams}, \& {Proctor}}]{Iskakov:2007:SSD_low_Pm}
{Iskakov}, A.~B., {Schekochihin}, A.~A., {Cowley}, S.~C., {McWilliams}, J.~C.,
  \& {Proctor}, M.~R.~E. 2007, Physical Review Letters, 98, 208501

\bibitem[{{Katsukawa} \& {Orozco Su{\'a}rez}(2012)}]{Katsukawa:2012:power}
{Katsukawa}, Y., \& {Orozco Su{\'a}rez}, D. 2012, \apj, 758, 139

\bibitem[{{Khomenko} {et~al.}(2003){Khomenko}, {Collados}, {Solanki}, {Lagg},
  \& {Trujillo Bueno}}]{Khomenko:2003:quiet_sun_infrared}
{Khomenko}, E.~V., {Collados}, M., {Solanki}, S.~K., {Lagg}, A., \& {Trujillo
  Bueno}, J. 2003, \aap, 408, 1115

\bibitem[{{Lites}(2011)}]{Lites:2011:hinode_ssd}
{Lites}, B.~W. 2011, \apj, 737, 52

\bibitem[{{Lites} {et~al.}(1996){Lites}, {Leka}, {Skumanich}, {Martinez
  Pillet}, \& {Shimizu}}]{Lites:1996:small_scale_field}
{Lites}, B.~W., {Leka}, K.~D., {Skumanich}, A., {Martinez Pillet}, V., \&
  {Shimizu}, T. 1996, \apj, 460, 1019

\bibitem[{{Lites} {et~al.}(2008){Lites}, {Kubo}, {Socas Navarro}, {Berger},
  {Frank}, {Shine}, {Tarbell}, {Title}, {Ichimoto}, {Katsukawa}, {Tsuneta},
  {Suematsu}, {Shimizu}, \& {Nagata}}]{Lites:etal:2008}
{Lites}, B.~W., {Kubo}, M., {Socas Navarro}, H., {et~al.} 2008, \apj, 672, 1237

\bibitem[{{Mart{\'{\i}}nez Gonz{\'a}lez} {et~al.}(2008){Mart{\'{\i}}nez
  Gonz{\'a}lez}, {Asensio Ramos}, {L{\'o}pez Ariste}, \& {Manso
  Sainz}}]{MartinezGonzales:2008:iso}
{Mart{\'{\i}}nez Gonz{\'a}lez}, M.~J., {Asensio Ramos}, A., {L{\'o}pez Ariste},
  A., \& {Manso Sainz}, R. 2008, \aap, 479, 229

\bibitem[{{Mart{\'{\i}}nez Pillet}(2013)}]{Mpillet:2013:SSRv}
{Mart{\'{\i}}nez Pillet}, V. 2013, \ssr, 178, 141

\bibitem[{{Nelson} {et~al.}(2013){Nelson}, {Brown}, {Brun}, {Miesch}, \&
  {Toomre}}]{Nelson:2013:Wreaths&Cycles}
{Nelson}, N.~J., {Brown}, B.~P., {Brun}, A.~S., {Miesch}, M.~S., \& {Toomre},
  J. 2013, \apj, 762, 73

\bibitem[{{Orozco Su{\'a}rez} \& {Bellot
  Rubio}(2012)}]{Orozco:Rubio:2012:QS_LP}
{Orozco Su{\'a}rez}, D., \& {Bellot Rubio}, L.~R. 2012, \apj, 751, 2

\bibitem[{{Orozco Su{\'a}rez} {et~al.}(2007){Orozco Su{\'a}rez}, {Bellot
  Rubio}, {del Toro Iniesta}, {Tsuneta}, {Lites}, {Ichimoto}, {Katsukawa},
  {Nagata}, {Shimizu}, {Shine}, {Suematsu}, {Tarbell}, \&
  {Title}}]{Orozco-Suarez:etal:2007}
{Orozco Su{\'a}rez}, D., {Bellot Rubio}, L.~R., {del Toro Iniesta}, J.~C.,
  {et~al.} 2007, \apjl, 670, L61

\bibitem[{{Petrovay} \& {Szakaly}(1993)}]{Petrovay:1993:SSD}
{Petrovay}, K., \& {Szakaly}, G. 1993, \aap, 274, 543

\bibitem[{{Pietarila Graham} {et~al.}(2010){Pietarila Graham}, {Cameron}, \&
  {Sch{\"u}ssler}}]{Pietarila-Graham:etal:2010:SSD}
{Pietarila Graham}, J., {Cameron}, R., \& {Sch{\"u}ssler}, M. 2010, \apj, 714,
  1606

\bibitem[{{Rempel}(2006)}]{Rempel:2006:dynamo}
{Rempel}, M. 2006, \apj, 647, 662

\bibitem[{{Rempel} {et~al.}(2009){Rempel}, {Sch{\"u}ssler}, \&
  {Kn{\"o}lker}}]{Rempel:etal:2009}
{Rempel}, M., {Sch{\"u}ssler}, M., \& {Kn{\"o}lker}, M. 2009, \apj, 691, 640

\bibitem[{{Rogers} {et~al.}(1996){Rogers}, {Swenson}, \&
  {Iglesias}}]{Rogers:opal:1996}
{Rogers}, F.~J., {Swenson}, F.~J., \& {Iglesias}, C.~A. 1996, \apj, 456, 902

\bibitem[{{S{\'a}nchez Almeida}(2003)}]{Almeida:2003:internetwork_sol_min}
{S{\'a}nchez Almeida}, J. 2003, \aap, 411, 615

\bibitem[{{Schekochihin} {et~al.}(2004){Schekochihin}, {Cowley}, {Taylor},
  {Maron}, \& {McWilliams}}]{Schekochihin:2004:SSD_high_Pm}
{Schekochihin}, A.~A., {Cowley}, S.~C., {Taylor}, S.~F., {Maron}, J.~L., \&
  {McWilliams}, J.~C. 2004, \apj, 612, 276

\bibitem[{{Schekochihin} {et~al.}(2007){Schekochihin}, {Iskakov}, {Cowley},
  {McWilliams}, {Proctor}, \& {Yousef}}]{Schekochihin:2007:SSD_low_PM}
{Schekochihin}, A.~A., {Iskakov}, A.~B., {Cowley}, S.~C., {et~al.} 2007, New
  Journal of Physics, 9, 300

\bibitem[{{Sch{\"u}ssler}(1990)}]{Schuessler:1990:IAUS}
{Sch{\"u}ssler}, M. 1990, in IAU Symposium, Vol. 138, Solar Photosphere:
  Structure, Convection, and Magnetic Fields, ed. J.~O. {Stenflo}, 161

\bibitem[{{Sch{\"u}ssler} \&
  {V{\"o}gler}(2008)}]{Schuessler:Voegler:2008:bhorz}
{Sch{\"u}ssler}, M., \& {V{\"o}gler}, A. 2008, \aap, 481, L5

\bibitem[{{Shchukina} \& {Trujillo Bueno}(2011)}]{Shchukina:2011:hanle_dynamo}
{Shchukina}, N., \& {Trujillo Bueno}, J. 2011, \apjl, 731, L21

\bibitem[{{Spruit}(1979)}]{Spruit:1979:convcollapse}
{Spruit}, H.~C. 1979, \solphys, 61, 363

\bibitem[{{Stein} {et~al.}(2003){Stein}, {Bercik}, \&
  {Nordlund}}]{Stein:2003:recirculation}
{Stein}, R.~F., {Bercik}, D., \& {Nordlund}, {\AA}. 2003, in Astronomical
  Society of the Pacific Conference Series, Vol. 286, Current Theoretical
  Models and Future High Resolution Solar Observations: Preparing for ATST, ed.
  A.~A. {Pevtsov} \& H.~{Uitenbroek}, 121

\bibitem[{{Stenflo}(2012)}]{Stenflo:2012:scaling}
{Stenflo}, J.~O. 2012, \aap, 541, A17

\bibitem[{{Stenflo}(2013)}]{Stenflo:2013:inc}
---. 2013, \aap, 555, A132

\bibitem[{{Tobias} {et~al.}(2011){Tobias}, {Cattaneo}, \&
  {Boldyrev}}]{Tobias:Cattaneao:Boldyrev:SSD_review}
{Tobias}, S.~M., {Cattaneo}, F., \& {Boldyrev}, S. 2011, ArXiv e-prints,
  arXiv:1103.3138

\bibitem[{{Trujillo Bueno} {et~al.}(2004){Trujillo Bueno}, {Shchukina}, \&
  {Asensio Ramos}}]{Trujillo:etal:2004}
{Trujillo Bueno}, J., {Shchukina}, N., \& {Asensio Ramos}, A. 2004, \nat, 430,
  326

\bibitem[{{V{\" o}gler} {et~al.}(2005){V{\" o}gler}, {Shelyag}, {Sch{\"
  u}ssler}, {Cattaneo}, {Emonet}, \& {Linde}}]{Voegler:etal:2005}
{V{\" o}gler}, A., {Shelyag}, S., {Sch{\" u}ssler}, M., {et~al.} 2005, \aap,
  429, 335

\bibitem[{{V{\"o}gler} \& {Sch{\"u}ssler}(2007)}]{Voegler:Schuessler:2007}
{V{\"o}gler}, A., \& {Sch{\"u}ssler}, M. 2007, \aap, 465, L43

\end{thebibliography}

\appendix
We present here the definitions of the transfer functions that are discussed in detail in Section 
\ref{sec:transfer}. We follow here mostly the derivation of \citet{Pietarila-Graham:etal:2010:SSD}, but use a 
slightly different definition for the transfer functions based on the momentum equation. We also use
a slightly different nomenclature, in the following a term $T_{XY}$ refers to an energy transfer to the
reservoir $X$ by a process $Y$. Here $X$ is either kinetic energy ($K$) or magnetic energy ($M$).
In the kinetic energy reservoir $Y$ refers to advection ($A$), pressure/buoyancy ($P$), and
Lorentz force ($L$), in the magnetic energy reservoir to advection ($A$), stretching ($S$), and
compression ($C$). In both reservoirs $D$ refers to contributions from numerical diffusivity.

The spectral magnetic energy is given by (here "$\widehat{\ldots}$" denotes the Fourier transform and "$^*$" the complex conjugate)
\begin{equation}
  E_M(k)=\frac{1}{8\pi}\widehat{\vec{B}}(k)\cdot\widehat{\vec{B}}^*(k)
\end{equation}
The time evolution of $E_M(k)$ is given by
\begin{equation}
  \frac{\partial}{\partial t}E_M(k)=\frac{1}{8\pi}\left(\widehat{\vec{B}}(k)\cdot\frac{\partial\widehat{\vec{B}}^*(k)}{\partial t}
    +\widehat{\vec{B}}^*(k)\cdot\frac{\partial\widehat{\vec{B}}(k)}{\partial t}\right)
\end{equation}
Using the induction equation we can write the time evolution of $E_M(k)$ as
\begin{equation}
  \frac{\partial}{\partial t}E_M(k)=T_{MA}(k)+T_{MS}(k)+T_{MC}(k)+T_{MD}(k)
\end{equation}
with the terms (here $c.c.$ refers to the complex conjugate expression)
\begin{eqnarray}
  T_{MA}(k)&=&-\frac{1}{8\pi}\widehat{\vec{B}}(k)\cdot\widehat{[\vec{v}\cdot\nabla \vec{B}]}^*(k)+c.c.\\
  T_{MS}(k)&=&\frac{1}{8\pi}\widehat{\vec{B}}(k)\cdot\widehat{[\vec{B}\cdot\nabla \vec{v}]}^*(k)+c.c.\\
  T_{MC}(k)&=&-\frac{1}{8\pi}\widehat{\vec{B}}(k)\cdot\widehat{[\vec{B}\nabla\cdot \vec{v}]}^*(k)+c.c.
\end{eqnarray}
Here the terms $T_{MA}$, $T_{MS}$, and $T_{MC}$ are energy transfers to magnetic energy through advection, shear 
and compression. $T_{MD}$ denotes the energy transfer due to numerical diffusivity according to the scheme described 
in Section \ref{sect:numdiff}.

We define the spectral kinetic energy through the expression
\begin{equation}
  E_K(k)=\frac{1}{2}\widehat{[\sqrt{\varrho}\vec{v}]}(k)\cdot\widehat{[\sqrt{\varrho}\vec{v}]}^*(k)
\end{equation}
The time evolution of $E_K(k)$ is given by
\begin{equation}
  \frac{\partial }{\partial t}E_K(k)=\frac{1}{2}\left(\widehat{[\sqrt{\varrho}\vec{v}]}(k)\cdot\frac{\partial
      \widehat{[\sqrt{\varrho}\vec{v}]}^*(k)}{\partial t}
    +\widehat{[\sqrt{\varrho}\vec{v}]}^*(k)\cdot\frac{\partial \widehat{[\sqrt{\varrho}\vec{v}]}(k)}{\partial t}\right)
\end{equation}
To derive the corresponding transfer functions we need an expression for the time evolution of the quantity $\sqrt{\varrho}v$, 
which follows from the momentum equation as
\begin{equation}
  \frac{\partial \sqrt{\varrho}\vec{v}}{\partial t}=\frac{1}{\sqrt{\varrho}}\left[-\nabla\cdot(\varrho\vec{v}\vec{v})
    -\frac{1}{2}\vec{v}\nabla\cdot(\varrho\vec{v})-\nabla P + \varrho\vec{g}+\frac{1}{4\pi}\nabla\cdot\left(\vec{B}\vec{B}-\frac{1}{2}\vec{I}\,B^2
      \right)\right]\;.
\end{equation}
This leads to
\begin{equation}
  \frac{\partial}{\partial t}E_K(k)=T_{KA}(k)+T_{KP}(k)++T_{KL}+T_{KD}(k)
\end{equation}
with the transfer functions:
\begin{eqnarray}
  T_{KA}(k)&=&-\frac{1}{2}\widehat{[\sqrt{\varrho}\vec{v}]}(k)\cdot\widehat{\left[\frac{\nabla\cdot(\varrho\vec{v}\vec{v})}
      {\sqrt{\varrho}}\right]}^*(k)-\frac{1}{4}\widehat{[\sqrt{\varrho}\vec{v}]}(k)\cdot\widehat{\left[
      \frac{\vec{v}\nabla\cdot(\varrho\vec{v})}{\sqrt{\varrho}}\right]}^*(k) +c.c.\\
  T_{KP}(k)&=&\frac{1}{2}\widehat{[\sqrt{\varrho}\vec{v}]}(k)\cdot\widehat{\left[\frac{-\nabla P
        + \varrho\vec{g}}{\sqrt{\varrho}}\right]}^*(k)+c.c.\\
  T_{KL}(k)&=&\frac{1}{8\pi}\widehat{[\sqrt{\varrho}\vec{v}]}(k)\cdot\widehat{\left[\frac{\nabla\cdot(\vec{B}\vec{B})}{\sqrt{\varrho}}
    \right]}^*(k) - \frac{1}{8\pi}\widehat{[\sqrt{\varrho}\vec{v}]}(k)\cdot\widehat{\left[\frac{\nabla B^2}
      {2\sqrt{\varrho}}\right]}^*(k)+c.c.
\end{eqnarray}
Here the terms $T_{KA}$ is the energy transfer within the kinetic energy reservoir by advection, $T_{KP}$ is the energy transfer from
pressure forces and buoyancy, and $T_{KL}$ is the energy transfer via Lorentz force. $T_{KD}$ denotes the energy transfer due to 
numerical diffusivity according to the scheme described in Section \ref{sect:numdiff}.

Similar to \citet{Pietarila-Graham:etal:2010:SSD} we use only Fourier transformations in the horizontal directions to compute
all transfer functions and average those over a certain height range in the vertical direction. The resulting
expressions do not account for the contributions from spectral energy transport in the vertical direction and as a consequence
do not necessarily balance each other completely.  

\end{document}